\documentclass[showpacs,showkeys,twocolumn,nofootinbib]{revtex4}

\usepackage{amsmath}
\usepackage{amssymb}
\usepackage{slashed}
\usepackage[dvips]{graphicx}
\usepackage{color}

\usepackage{verbatim} 
\usepackage[sans]{dsfont}

\newcommand{\bra}[1]{\langle#1\vert}
\newcommand{\ket}[1]{\vert#1\rangle}
\newcommand{\Tr}{\mathop{\rm Tr}\nolimits}
\newcommand{\e}{\ensuremath{\mathrm{e}}}
\renewcommand{\d}{\ensuremath{\mathrm{d}}}
\newcommand{\hc}{\ensuremath{\mathrm{h.c.}}}
\newcommand{\C}{\ensuremath{c}}
\newcommand{\T}{\ensuremath{\mathrm{T}}}
\newcommand{\F}{\ensuremath{\mathrm{F}}} 

\newcommand{\SU}[1]{\ensuremath{\mathrm{SU}(#1)}}

\newcommand{\OO}[1]{\ensuremath{\mathrm{O}(#1)}}
\newcommand{\U}[1]{\ensuremath{\mathrm{U}(#1)}}

\renewcommand{\d}{\ensuremath{\mathrm{d}}}

\begin{document}
\def\im{\mathrm{i}}
\def \openone {\leavevmode\hbox{\small1\kern-3.0pt\normalsize1}}
\def\beginm{\left(\begin{array}}
\def\endm{\end{array}\right)}

\title{Sterile Particles from the Flavor Gauge Model of Masses}

\author{Adam Smetana}
\email{smetana@ujf.cas.cz}
\affiliation{Department of Theoretical Physics, Nuclear Physics Institute, \v Re\v z (Prague), Czech Republic}

\begin{abstract}
The existence of right-handed neutrinos follows from theoretical consistence of the recently suggested electroweak symmetry breaking model, based on dynamical flavor gauge symmetry breaking. Only finite number of versions of the model exists. They differ by the number and the flavor structure of the right-handed neutrinos. We choose for inspection one of them, the non-minimal version with right-handed neutrinos in sextet flavor representation, and at some points we compare it with the minimal version. We show that a Majorana pairing of the sextet right-handed neutrinos is responsible for the flavor symmetry breaking, and the seesaw pattern of the neutrino mass matrix naturally arises. The dynamically generated neutrino mass matrix spontaneously breaks the lepton number and the chiral sterility symmetry of the right-handed neutrino sector. As a result spectrum of majorons, neutrino composites, manifests. We study main characteristics of both massive sterile neutrinos and majorons which show their relevance as dark matter candidates.
\end{abstract}



\maketitle

\section{Introduction}

With recent significant improvement of quality of cosmological,
astrophysical and neutrino observations, the neutrino sector of
particle spectrum is just becoming increasingly powerful tool to
discriminate among various models of the electroweak symmetry
breaking. Recently suggested model \cite{Hosek:2009ys,Hosek:NagoyaProceeding,Benes:2011gi}
of dynamically generated masses turns out to be extremely difficult
to be approved or disproved through direct computation of its
mass spectrum. Nevertheless, the model provides clear and clean
predictions about the structure of the right-handed neutrino sector,
about its global symmetries, and about majorons, the composite Nambu--Goldstone
scalars, the consequences of the spontaneous breaking of the global symmetries.

The right-handed neutrinos are often proposed to exist for their power
to explain straightforwardly the observed neutrino masses and, especially,
to explain why the neutrinos are so light via the see-saw mechanism
\cite{GellMann:1980vs,Mohapatra:1979ia,Yanagida:1979as}. Since they are
Standard Model singlets they do not produce color and electroweak gauge
anomaly. Therefore, if they are not charged with respect to some new gauge
force, their number is not constrained. In models where the family or flavor
$\SU{3}_\F$ index is gauged \cite{Wilczek:1978xi,Ong:1978tq,Davidson:1979wt,Chakrabarti:1979vy,Yanagida:1979gs,Yanagida:1980xy,Berezhiani:1990wn,Nagoshi:1990wk}
the number of all fields that feel the new flavor gauge force has to be
balanced so that the flavor gauge anomaly cancels. The overall contribution
of observed electroweakly charged fermions to the flavor gauge anomaly does
not vanish. Therefore additional fields, the chromodynamically and electroweakly
neutral right-handed neutrinos, are needed in a specific number \cite{Kribs:2003jn}.

The flavor gauge model studied in this paper intends not to
postulate any further new dynamics and leaves whole responsibility
for the electroweak symmetry breaking on the gauge flavor
$\SU{3}_\mathrm{F}$ dynamics. In order to make sense the gauge
flavor dynamics is strong, asymptotically free, self-breaking, and
non-confining, i.e., non-vector-like\footnote{Non-vector-like gauge
theory arises from gauging not only vector currents, like in QCD,
but also axial-vector currents.}.

The flavor gauge symmetry breaking, the cause for the mass generation, is so far
only assumed in the model. It is supposed to be achieved neither by the vacuum
expectation value of a scalar field, nor by the confining strong gauge dynamics,
but it is the strong flavor gauge dynamics itself that self-breaks. As there is
no similar effect observed in nature, as the effort to put the chiral theories
on a lattice fails, and as the solution of the equations of the model are painfully
unattainable, it is not clear whether the self-breaking mechanism is
possible at all. In general, it is not clear whether a chiral gauge dynamics
alone could dynamically generate self-energies by which it breaks its own gauge
symmetry. On the other hand this scenario pioneered by \cite{Eichten:1974et} has
not been disqualified yet.

Fortunately, the gauge symmetries tights the model so much that there is no room for
fine-tuning and firm predictions arise. Theoretical consistence
of the flavor gauge model predicts the number of right-handed neutrinos with only
little ambiguity. The model admits right-handed neutrinos only in selected flavor
representation settings whose number is finite and not large. Throughout the paper
we bring more or less heuristic arguments for that there is only
one right-handed neutrino setting that defines phenomenologically viable and
preferred version of the model.

The preferred version of the model is non-minimal in the sense
that it contains the right-handed neutrinos in a flavor sextet
representation. It provides appealing features: (i) It is chiral,
i.e., the only mass scale comes from the dimensional transmutation
of the running flavor coupling constant. (ii) It is essentially
non-vector-like. (iii) The sextet right-handed neutrino Majorana
pairing leads naturally to the see-saw pattern of the neutrino
mass matrix. (iv) It provides light sterile neutrinos, in addition
to the three electroweak neutrinos what can be of particular
interest with respect to the dark matter
\cite{Nieuwenhuizen:2008pf,Kusenko:2009up,Bezrukov:2009th}. (v)
The dynamically generated neutrino mass matrix breaks
spontaneously both the lepton number and the sterility symmetry
$G_\mathrm{S}$, the accidental global symmetry of the right-handed
neutrino sector. As a result, numerous Nambu--Goldstone
neutrino-composites, called majorons
\cite{Chikashige:1980ui,Schechter:1981cv}, appear in the spectrum.
The existence of majorons, especially the standard majoron, is a
rigid prediction present in all versions of the model, while they
do \emph{not} present a phenomenological danger in form of
long-range force \cite{Gelmini:1982zz}.

The aim of the paper is to point out aspects of the sterile sector
of the non-minimal version of the flavor gauge model, and document
both theoretically and phenomenologically why others are not so preferable.

The paper is organized as follows. In section~\ref{secII} we investigate
the right-handed neutrino structure of the flavor gauge model: After a brief
recapitulation of the model we summarize all viable versions of the model.
We argue why we choose the non-minimal version for the
analysis in the rest of the paper. In section~\ref{secIII} we write the
part of the model Lagrangian relevant for the neutrino sector. We analyze
its global sterility symmetry $G_\mathrm{S}$. In section~\ref{secIV} we
apply the model idea to generate neutrino masses, and discuss the flavor
symmetry breaking. In the non-minimal version in contrast to the minimal
one, the privileged role of right-handed neutrinos is recognized: It is
their Majorana pairing which triggers the flavor symmetry breaking. In
section~\ref{secV} we analyze the majoron spectrum arising from spontaneous
lepton number and sterility symmetry breaking. In section~\ref{secVI} we
conclude.

\section{Right-handed neutrino fields}\label{secII}

The quantum flavor gauge dynamics of the model requires the
existence of the right-handed neutrinos and restricts severely
their number. As a main result of this section we list the finite
number of all acceptable flavor settings that are anomaly and
asymptotically free, and do not provide the perturbative infrared
fixed point. These three properties are necessary for a viability
of the model. Later, we rather heuristically argue that some
settings are more preferable than others.

First we briefly recapitulate the model.

\subsection{Flavor gauge model}
The basis of the model is that the chiral electroweak symmetry is broken dynamically
by chirality changing fermion self-energies $\Sigma(p^2)$ generated by the
strong flavor dynamics. The flavor structure of the self-energies $\Sigma(p^2)$
is crucial for it should reflect the hierarchical pattern of fermion masses.

The model is defined by the flavor setting of electroweakly
charged Weyl fermions. There are two distinct cases, case I and
case II, see Tab.~\ref{FermionSetting}. The ultimate
discrimination among them can be made after the successful
solution of mass equations is found, or after full structure of
the neutrino sector is revealed.

\begin{table}[t]
\begin{tabular}{lccccc|c}
 & $q_\mathrm{L}$ & $u_\mathrm{R}$ & $d_\mathrm{R}$ & $\ell_\mathrm{L}$ & $e_\mathrm{R}$ & $N$  \\
\hline
\hline
case I  & $\mathbf{3}$ & $\mathbf{3}$ & $\overline{\mathbf{3}}$ & $\overline{\mathbf{3}}$ & $\mathbf{3}$ & $3$  \\
case II & $\mathbf{3}$ & $\mathbf{3}$ & $\overline{\mathbf{3}}$ & $\overline{\mathbf{3}}$ & $\overline{\mathbf{3}}$ & $5$ \\
\hline
\hline
\end{tabular}
\caption{\small Two possible flavor settings of electroweakly charged fermions. The number $N$ tells how many flavor triplets are necessary to cancel the flavor gauge anomaly. The notation is obvious: $q_\mathrm{L}=(u_\mathrm{L},d_\mathrm{L})^\T$, $\ell_\mathrm{L}=(\nu_\mathrm{L},e_\mathrm{L})^\T$, $u=(u,c,t)$, $d=(d,s,b)$, $\nu=(\nu_e,\nu_\mu,\nu_\tau)$, and $e=(e,\mu,\tau)$. }
\label{FermionSetting}
\end{table}

The purpose of this setting is to distinguish self-energy matrices
for fermions of various charges, as generally\footnote{The flavor structure of self-energies should be understood via the corresponding mass terms $\overline{f_\mathrm{R}(\mathbf{r}^\prime)}\Sigma^{\mathbf{r}^\prime\hspace{-3pt}\times\bar{\mathbf{r}}}f_\mathrm{L}(\mathbf{r})$ where $\mathbf{r}$ and $\mathbf{r}^\prime$ are flavor representations of $f_\mathrm{L}$ and $f_\mathrm{R}$, respectively. }
$\Sigma^{\mathbf{3}\times\mathbf{3}}\neq\Sigma^{\overline{\mathbf{3}}\times\mathbf{3}}\neq
\Sigma^{\mathbf{3}\times\overline{\mathbf{3}}}\neq\Sigma^{\overline{\mathbf{3}}\times\overline{\mathbf{3}}}$.
(This idea was also pursued in the class of extended technicolor
models \cite{Appelquist:2003hn}.) In order to achieve the
exclusivity of the $u$-type quarks, whose observed mass spectrum
is significantly heavier, \emph{we prefer the case I}. Their self-energy
is of type $\Sigma^{\overline{\mathbf{3}}\times\mathbf{3}}$ as distinct from
$d$-type and $e$-type fermion self-energies which are of type
$\Sigma^{\overline{\mathbf{3}}\times\overline{\mathbf{3}}}$ and
$\Sigma^{\mathbf{3}\times\mathbf{3}}$. Neutrino
self-energies are distinguished from others by their Majorana
components and by possible higher flavor representation settings
of right-handed neutrinos. Due to the flavor setting the mass
hierarchy among different charges can be achieved. The mass
hierarchy among generations then has completely different origin.
It follows from the fact that the flavor symmetry is completely broken
providing distinct eigenvalues of the self-energies.

The characteristic fermion flavor setting plays also another important
role. It makes the flavor gauge dynamics non-vector-like, what distinguishes
it from QCD and makes it non-confining.

As well as the QCD, the flavor gauge dynamics is asymptotically
free, i.e., the effective flavor gauge coupling constant $\bar
h(q^2)$ in perturbative regime runs according to
\begin{equation}
\frac{\bar h^2(q^2)}{4\pi}=\frac{4\pi}{(11-\frac{1}{3}N^\mathrm{EW}-\frac{2}{3}\eta_{\nu_\mathrm{R}})\ln{q^2/\Lambda^{2}_\mathrm{F}}}\,,
\end{equation}
where $N^{\mathrm{EW}}=15$ is the number of electroweakly charged
flavor triplets and $\eta_{\nu_\mathrm{R}}$ is the right-handed
neutrino contribution to the coefficient of the flavor
$\beta$-function. Well above the scale of the flavor gauge
dynamics, $\Lambda_\mathrm{F}$, everything is weakly coupled and
symmetric. Decreasing the energy scale the effective flavor gauge
coupling increases till it surpasses its critical value at the
energy scale around $\Lambda_\mathrm{F}$. Because of its
non-vector-like nature the flavor symmetry itself does not survive
anymore and is spontaneously broken
\cite{Eichten:1974et,Pagels:1979ai}. The flavor gauge bosons
acquire masses of order of the flavor symmetry breaking scale
$\Lambda_\mathrm{F}$. (For details see
\cite{Benes:2011gi}.)

The flavor gauge bosons have to be enormously heavy in order to suppress
the processes with flavor changing neutral currents, giving a lower bound
for their mass to be more than $10^6\,\mathrm{GeV}$ \cite{Eichten:1979ah}. But in order to make
the axion, which is naturally present in the model, invisible it is better to assume
that the quark self-energies are formed at the scale in the so called axion
window $10^{10}\,\mathrm{GeV}<\Lambda_\mathrm{F}<10^{12}\,\mathrm{GeV}$
\cite{Raffelt:2006rj}. We will see that in the non-minimal versions of the
model the right-handed neutrino Majorana self-energy should be generated
at even much higher scale.

The `would-be' Nambu--Goldstone bosons of the broken electroweak
symmetry, which are composites of fermions, manifest themselves as the
longitudinal components of the electroweak gauge bosons, producing their
masses. The electroweak gauge boson masses are therefore directly, though
non-trivially linked to the masses of electroweakly charged fermions.
Therefore we expect $M_{W,Z}$ being proportional rather to $m_t$ and not
to some electroweak scale $\Lambda_\mathrm{EW}$, which in fact does not
exist in this model.

\subsection{Constraints on the number of right-handed neutrino fields}

\subsubsection{Anomaly freedom}
The model would suffer from the flavor gauge anomaly unless the
proper number of right-handed neutrino fields is added into the
model. They are needed to compensate the non-zero flavor anomaly
contribution of electroweakly charged fermions. In
Tab.~\ref{FermionSetting} the number $N$ indicates that 3 (5)
additional triplets of right-handed neutrinos make the flavor
gauge dynamics anomaly free.

Adding of triplets is not the only possibility. Specially balanced
settings including higher representations, sextet, octet, or
decuplet, etc., lead to the anomaly free models too. Constructing
the non-minimal versions of the model, notice that a pair of complex
multiplet and its conjugate, as well as real representation multiplet
do not contribute to the anomaly.

\subsubsection{Asymptotic freedom}
On the other hand, we should not add too many right-handed neutrinos
in order not to destroy the asymptotic freedom of the flavor
dynamics. Within the one-loop approximation of the
$\beta$-function, the $\eta_{\nu_\mathrm{R}}$ coefficient is
constrained as
\begin{eqnarray}
\eta_{\nu_\mathrm{R}} & \equiv & 1/2N^{\nu_\mathrm{R}}_3+5/2N^{\nu_\mathrm{R}}_6+ \nonumber\\
& & 3N^{\nu_\mathrm{R}}_8+15/2N^{\nu_\mathrm{R}}_{10}+\ldots<9 \,,\label{AF_inequality}
\end{eqnarray}
where $N^{\nu_\mathrm{R}}_r$ is the number of right-handed neutrino
multiplets of a given representation $\mathbf{r}$ and $\overline{\mathbf{r}}$.
The inequality \eqref{AF_inequality} leaves us to combine only lower
dimensional multiplets, $\mathbf{3}$, $\overline{\mathbf{3}}$,
$\mathbf{6}$, $\overline{\mathbf{6}}$, and $\mathbf{8}$.

\subsubsection{Absence of the perturbative infrared fixed point}
Even more stringent limit comes from demand not to produce too
small, i.e., sub-critical, pertubative infrared fixed point, say
$\alpha^{*}_{\mathrm{F,\,IR}}<0.5$, where
$\alpha_\mathrm{F}\equiv\frac{h^2}{4\pi}$. It would leave the
system in the chirally symmetric phase and prevent the whole
symmetry breaking mechanism.

We choose the discriminating value of
$\alpha^{*}_{\mathrm{F,\,IR}}$ being $0.5$ quite arbitrarily but
motivated by QCD running coupling constant which is measured
(still being in a perturbative regime) at the scale
$1.7\,\mathrm{GeV}\gtrsim\Lambda_{\mathrm{QCD}}$ having the value
$\alpha_\mathrm{s}(1.7\,\mathrm{GeV})\thickapprox0.35$
\cite{Nakamura:2010zzi}.

A zero of the two-loop $\beta$-function \eqref{beta2} gives an
estimate of the perturbative infrared fixed point
\begin{equation}
\alpha^{*}_{\mathrm{F,\,IR}}=-4\pi\frac{-18+N^{\nu_\mathrm{R}}_3+5N^{\nu_\mathrm{R}}_6+6N^{\nu_\mathrm{R}}_8}
                                       {-21+19N^{\nu_\mathrm{R}}_3+125N^{\nu_\mathrm{R}}_6+144N^{\nu_\mathrm{R}}_8} \,.
\end{equation}

\subsubsection{Chirality and non-vector-like nature}
Putting all together we get only few possible right-handed
neutrino flavor settings defining still viable models. We list
them in Tab.~\ref{NUsettings}. The models fall into various
classes according to two criteria, their chirality and their
approximate vector-like nature.

\begin{table}[t]
\begin{tabular}{l|l|c|c}
\hline
\hline
&       &        & approx. \\
& $\nu_\mathrm{R}$ representation setting & chiral & vector-like \\
&       &        & around $\Lambda_\mathrm{F}$ \\
\hline
case I  & $3\times\mathbf{3}$ & \textbf{yes} & yes \\
        & $3\times\mathbf{3},\ 1\times(\mathbf{3},\overline{\mathbf{3}})$ & no & yes \\
        & $3\times\mathbf{3},\ 2\times(\mathbf{3},\overline{\mathbf{3}})$ & no & yes \\
        & $3\times\mathbf{3},\ 3\times(\mathbf{3},\overline{\mathbf{3}})$ & no & yes \\
        & $1\times\mathbf{6},\ 4\times\overline{\mathbf{3}}$ & \textbf{yes} & \textbf{no} \\
        & $1\times\mathbf{8},\ 3\times\mathbf{3}$ & no & \textbf{no} \\
\hline
case II & $5\times\mathbf{3}$ & \textbf{yes} & yes \\
        & $5\times\mathbf{3},\ 1\times(\mathbf{3},\overline{\mathbf{3}})$ & no & yes \\
        & $5\times\mathbf{3},\ 2\times(\mathbf{3},\overline{\mathbf{3}})$ & no & yes \\
        & $1\times\mathbf{6},\ 2\times\overline{\mathbf{3}}$ & \textbf{yes} & \textbf{no} \\
        & $1\times\mathbf{6},\ 2\times\mathbf{3},\ 1\times(\mathbf{3},\overline{\mathbf{3}})$ & no & \textbf{no} \\
\hline \hline
\end{tabular}
\caption{\small All viable versions of the flavor gauge model. }
\label{NUsettings}
\end{table}

The models containing right-handed neutrinos in both $\mathbf{3}$,
and $\overline{\mathbf{3}}$, or in $\mathbf{8}$, allow the gauge
invariant hard Majorana mass term. Therefore they are non-chiral
possessing a hard Majorana mass parameter. The origin of such mass
parameter is not explained by the model and it would have been
assumed to follow from yet another dynamics operating at higher
energy scale. In this sense the chiral models appear to be more
complete and more fundamental.

From the high energy (around $\Lambda_\mathrm{F}$) perspective,
the versions of the model that contain only $\mathbf{3}$, or
$\overline{\mathbf{3}}$ are approximately vector-like with small
non-vector-like perturbation given by the Standard Model gauge
dynamics. In that case the dynamics resembles the dynamics of QCD
and presumably prefers pairing in the
$\mathbf{3}\times\overline{\mathbf{3}}$ that does not ensure the
flavor symmetry breaking. The flavor breaking fermion
self-energies are then only believed to be energetically more
favorable than the flavor preserving ones. On the other hand, the
versions of the model that contain right-handed neutrinos in
higher representation $\mathbf{6}$, are essentially non-vector-like
and prefer right-handed neutrino pairing in the Majorana channels
$\mathbf{6}\times\overline{\mathbf{3}}$, or
$\mathbf{6}\times\mathbf{6}$, that certainly break the flavor
symmetry.

\begin{center}*\end{center}

The \emph{minimal} version with three
right-handed neutrino triplets denoted by (333) was analyzed in
the paper \cite{Benes:2011gi}. In this paper we will pursue the
\emph{non-minimal} version, the only case I version which is both
non-vector-like and chiral. Its right-handed neutrino setting is
$(\mathbf{6},\overline{\mathbf{3}},\overline{\mathbf{3}},\overline{\mathbf{3}},\overline{\mathbf{3}})$,
and we will denote it by (63333).

\section{Neutrino Lagrangian and its symmetries}\label{secIII}

In this section we define the non-minimal and preferred versions
of the flavor gauge model with the triplet right-handed electron
and with four right-handed neutrino anti-triplet and one
right-handed neutrino sextet by writing the Lagrangian of their
neutrino sector. Next we identify its sterility symmetry
$G_\mathrm{S}$.

\subsection{Lagrangian of neutrino sector}

The Lagrangian describing the neutrino flavor gauge dynamics is given by
\begin{eqnarray}
{\cal L} & = & -\frac{1}{4}F_{\mu\nu}^aF^{\mu\nu a}+{\cal L}_\nu \,,\\
\label{L}
{\cal L}_\nu & = &
\overline{\nu_\mathrm{L}}\gamma^\mu(\im\partial_\mu-hC^{a}_\mu T^{a*})\nu_\mathrm{L} \\
& &
+\sum_\mathbf{r}\overline{\nu_{\mathrm{R}\mathbf{r}}}\gamma^\mu(\im\partial_\mu+hC^{a}_\mu
T^{a}_\mathbf{r})\nu_{\mathrm{R}\mathbf{r}} \,,\nonumber
\end{eqnarray}
where the field strength tensor of flavor gauge bosons $C^{a}_\mu$
is given by $F_{\mu\nu}^a=\partial_\mu C^{a}_\nu-\partial_\nu
C^{a}_\mu+hf^{abc}C^{b}_\mu C^{\C}_\nu$. $T_{\mathbf{r}}^{a}$ are
$\SU{3}_\mathrm{F}$ generators for a representation $\mathbf{r}$
of the right-handed neutrino multiplet.\footnote{If the index
$\mathbf{r}$ is not used we mean the generators for the
fundamental triplet representation, given by the Gell-Mann
matrices $T^{a}=\frac{1}{2}\lambda^a$.}
The sum runs over one sextet with $T_{\mathbf{6}}^{a}$ and four
anti-triplets with
$T_{\overline{\mathbf{3}}}^{a}=-[T_{\mathbf{3}}^{a}]^*=-\frac{1}{2}\lambda^{*a}$.

This non-minimal version is chiral, i.e., it does not allow the
Majorana mass term,
$-\frac{1}{2}\overline{\nu^{\C}_{\mathrm{R}}}M\nu_{\mathrm{R}}$,
relevant for the non-chiral models.

\subsection{Global symmetries of neutrino sector}
\label{nuSymmetries}
Additionally to the global symmetries of the electroweakly
charged fermion sector of the model
\begin{equation}\label{GlobalSymmetriesEW}
\U{1}_\mathrm{B}\times\U{1}_\mathrm{L_{EW}}\times\U{1}_{\mathrm{B}_5}\times\U{1}_{\mathrm{L}_5} \,,
\end{equation}
(for detailed analysis of the global symmetries see
\cite{Benes:2011gi}) the sterile sector provides another global
symmetry of the classical Lagrangian \eqref{L}: a large sterility
symmetry $G_\mathrm{S}$, with both Abelian and non-Abelian
components. It is not ordered by anyone and comes out
accidentally.

The electroweak lepton number,
$L_\mathrm{EW}$,\footnote{$L_\mathrm{EW}$ denotes the lepton
number counting the electroweakly charged leptons, $e$,
$\nu_\mathrm{L}$, and \emph{not} the right-handed neutrinos
$\nu_\mathrm{R}$.} is defined by its current
\begin{equation}\label{LSM}
\mathcal{J}^{\mu}_{L_\mathrm{EW}}  =  \overline{e_\mathrm{L}}\gamma^\mu e_\mathrm{L}+\overline{e_\mathrm{R}}\gamma^\mu e_\mathrm{R}+\overline{\nu_\mathrm{L}}\gamma^\mu \nu_\mathrm{L} \,.
\end{equation}
As well as an Abelian part of the sterility symmetry (see below),
it is broken heavily by the flavor instanton effects due to the
flavor anomaly
\begin{equation}
\partial_\mu \mathcal{J}^{\mu}_{\mathrm{L}_\mathrm{EW}}=-\frac{h^2}{32\pi^2}F_{\alpha\beta a}\tilde{F}^{\alpha\beta}_a \,.
\end{equation}
(We neglect the electroweak anomaly.) Nevertheless one can always
find some linear combinations of the electroweak lepton number and the sterility symmetry
which are flavor anomaly free. One of them plays a role of the conserved lepton number $L$.

The setting of the right-handed neutrinos defines manifestly
chiral model. The chirality provides quite large accidental
sterility symmetry $G_\mathrm{S}$ of the right-handed neutrino
sector. The sterility symmetries are
\begin{equation}
G_\mathrm{S} =
\U{1}_{\mathrm{S}_6}\times\U{1}_{\mathrm{S}_3}\times\SU{4}_\mathrm{S}
\,.
\end{equation}
The corresponding Noether currents are
\begin{subequations}
\begin{eqnarray}
\mathcal{J}^{\mu}_{\mathrm{S}_6} & = & \overline{\nu_{\mathrm{R}\mathbf{6}}}\gamma^\mu\nu_{\mathrm{R}\mathbf{6}}=\Tr\overline{\xi_{\mathrm{R}}}\gamma^\mu\xi_{\mathrm{R}} \,; \\
\mathcal{J}^{\mu}_{\mathrm{S}_3} & = & \frac{1}{4}\overline{\zeta_{\mathrm{R}}^n}\gamma^\mu\zeta_{\mathrm{R}}^n \,;\\
\mathcal{J}^{\mu}_{\mathrm{S},i} & = &
\overline{\zeta_{\mathrm{R}}^n}\left[S_i\right]^{nm}\gamma^\mu\zeta_{\mathrm{R}}^m
\,,
\end{eqnarray}
\end{subequations}
where the summation over the flavor index is suppressed. The
indices $n,\,m=1,..,4$ run over four right-handed neutrino
anti-triplets. Matrices $S_i$, $i=1,..,15$, are
$\SU{4}_\mathrm{S}$ generators. We denote sextet right-handed
neutrinos as $\xi_\mathrm{R}$ and anti-triplet right-handed
neutrinos as $\zeta_\mathrm{R}$:
\begin{subequations}\label{RHnu}
\begin{eqnarray}
\xi_{\mathrm{R}} & \equiv & T^{\iota}_{\mathrm{(sym.)}}\nu^{\iota}_{\mathrm{R}\mathbf{6}} \,,\\
\zeta^{n}_{\mathrm{R}} & \equiv &
\nu_{\mathrm{R}\overline{\mathbf{3}}}^n \,,
\end{eqnarray}
\end{subequations}
where $\iota=1,..,6$ is the flavor index. The six symmetric $3\times3$
matrices $T^{\iota}_{\mathrm{(sym.)}}$ are
$\openone,\,\frac{1}{2}\lambda_1,\,\frac{1}{2}\lambda_3,\,\frac{1}{2}\lambda_4,\,\frac{1}{2}\lambda_6,\,\frac{1}{2}\lambda_8$.

The trace-full Abelian symmetries do not survive the incorporation
of the quantum effects. They are broken by anomalies what is
expressed by the non-vanishing four-divergencies of their
currents.\footnote{We ignore here the flavor anomalies of the
charged fermion currents corresponding to
\eqref{GlobalSymmetriesEW}. Their flavor anomalies would otherwise
provide some charged fermion component of the heavy sterile
majoron, see later. Ignoring their flavor anomalies allows us to
treat the heavy sterile majoron as a neutrino and flavor gauge
boson composite only.}
\begin{subequations}\label{anomaly}
\begin{eqnarray}
\partial_\mu \mathcal{J}^{\mu}_{\mathrm{S}_3-\mathrm{S}_6} & = & 0 \,;\\
\partial_\mu \mathcal{J}^{\mu}_{\mathrm{S},i} & = & 0 \,;\\
\partial_\mu
\mathcal{J}^{\mu}_{\mathrm{S}_3+\mathrm{S}_6} & = &
\frac{h^2}{16\pi^2}F_{\alpha\beta a}\tilde{F}^{\alpha\beta}_a \,.
\end{eqnarray}
\end{subequations}
Both $\mathcal{J}^{\mu}_{\mathrm{S}_3}$ and
$\mathcal{J}^{\mu}_{\mathrm{S}_6}$ are broken by the anomaly
individually, but their combination
$\mathcal{J}^{\mu}_{\mathrm{S}_3-\mathrm{S}_6}$ is exactly
conserved, while the orthogonal combination
$\mathcal{J}^{\mu}_{\mathrm{S}_3+\mathrm{S}_6}$ is not.

The conserved anomaly free lepton number $L$ and its current $\mathcal{J}^{\mu}_{L}$ are given as a linear combination
\begin{subequations}\label{Lepton_number}
\begin{eqnarray}
L & = & L_\mathrm{EW}+\left(a S_3+(1-a)S_6\right) \,, \\
\mathcal{J}^{\mu}_{L} & = & \mathcal{J}^{\mu}_{L_\mathrm{EW}}+\left(a \mathcal{J}^{\mu}_{S_3}+(1-a)\mathcal{J}^{\mu}_{S_6}\right) \,,
\end{eqnarray}
\end{subequations}
where the real coefficient $a$ is arbitrary.

\section{Massive neutrinos}\label{secIV}
Within the model, the neutrinos as well as all other fermions acquire masses due to the
strong flavor dynamics. In this section we describe the neutrino mass generation.
We continue with a discussion of the flavor symmetry breaking and by a neutrino phenomenology.

\subsection{Neutrino mass generation}

To treat the most general neutrino masses of both Majorana and
Dirac types in compact form, we introduce the neutrino multispinor
$n$ in the Nambu--Gorkov formalism
\begin{equation}\label{NGmultiplet}
n=\beginm{c} \nu_\mathrm{L}+(\nu_{\mathrm{L}})^\C \\ \nu_{\mathrm{R}\overline{\mathbf{3}}}^1+(\nu_{\mathrm{R}\overline{\mathbf{3}}}^1)^\C \\ \nu_{\mathrm{R}\overline{\mathbf{3}}}^2+(\nu_{\mathrm{R}\overline{\mathbf{3}}}^2)^\C \\
 \nu_{\mathrm{R}\overline{\mathbf{3}}}^3+(\nu_{\mathrm{R}\overline{\mathbf{3}}}^3)^\C \\ \nu_{\mathrm{R}\overline{\mathbf{3}}}^4+(\nu_{\mathrm{R}\overline{\mathbf{3}}}^4)^\C \\ \nu_{\mathrm{R}\mathbf{6}}+(\nu_{\mathrm{R}\mathbf{6}})^\C   \endm
\,,
\end{equation}
where the flavor indices are suppressed.

The Lagrangian \eqref{L} is then rewritten as
\begin{equation}
{\cal L}_\nu = \frac{1}{2}\bar{n}\gamma^\mu(\im\partial_\mu+hC^{a}_\mu t^{a})n \,,
\end{equation}
where the flavor generators $t^a$ in multi-component space are given in
\eqref{NGflavorGenerator}.

The chiral invariance underlying the gauge dynamics forbids to write the neutrino mass term
directly into the Lagrangian. The neutrino masses arise as poles of the full propagator\footnote{Here, we neglect
the wave function renormalization.} $S(p)\equiv[\slashed{p}-\mathbf{\Sigma}(p^2)]^{-1}$,
thus as solutions of the equation
\begin{equation}\label{detSigma}
\det\left(p^2-\mathbf{\Sigma}(p^2)\mathbf{\Sigma}^\dag(p^2)\right)=0\,.
\end{equation}
The neutrino self-energy $\mathbf{\Sigma}(p^2)$ is given as
\begin{equation}
\mathbf{\Sigma}(p^2) = \Sigma(p^2)P_\mathrm{L}+\Sigma^\dag(p^2)P_\mathrm{R} \,,
\end{equation}
where the \emph{symmetric} $21\times 21$ matrix $\Sigma(p^2)$ can
be written block-wise as
\begin{equation}
\Sigma=\beginm{cc}
    \Sigma_\mathrm{L} & \Sigma_\mathrm{D} \\ \Sigma_{\mathrm{D}}^\T & \Sigma_\mathrm{R} \endm
\end{equation}
or in more detail
\begin{equation}\label{NGselfenergy}
\Sigma=\beginm{c|c|c}
    L^{\mathbf{3}\times\mathbf{3}} & \ \ \ \ D^{\overline{\mathbf{3}}\times\mathbf{3}}_n \ \ \ \ &
                                                            D^{\mathbf{6}\times\mathbf{3}} \\ \hline
    D^{\mathbf{3}\times\overline{\mathbf{3}}}_m & R^{\overline{\mathbf{3}}\times\overline{\mathbf{3}}}_{mn} &
                    R^{\mathbf{6}\times\overline{\mathbf{3}}}_n \vphantom{\begin{array}{c}\ \\ \ \end{array}}\\ \hline
    D^{\mathbf{3}\times\mathbf{6}} & R^{\overline{\mathbf{3}}\times\mathbf{6}}_m & R^{\mathbf{6}\times\mathbf{6}} \\ \endm \,.
\end{equation}
By definition the self-energy matrix is symmetrical: the diagonal blocks, $L^{\mathbf{3}\times\mathbf{3}}$, $R^{\overline{\mathbf{3}}\times\overline{\mathbf{3}}}$ and $R^{\mathbf{6}\times\mathbf{6}}$ are symmetrical matrices, and $D^{\mathbf{3}\times\overline{\mathbf{3}}}_n=[D^{\overline{\mathbf{3}}\times\mathbf{3}}_n]^\T$, $D^{\mathbf{3}\times\mathbf{6}}=[D^{\mathbf{6}\times\mathbf{3}}]^\T$ and $R^{\overline{\mathbf{3}}\times\mathbf{6}}_n=[R^{\mathbf{6}\times\overline{\mathbf{3}}}_n]^\T$.

In the approximation of the truncated Schwinger--Dyson equation
with the wave function renormalization omitted the self-energy
is subject of the equation\footnote{We use the short-hand notation for integration $\int_k\equiv\int\frac{\d^4k}{(2\pi)^4}$.}
\begin{equation}\label{SDE}
\Sigma(p^2)  =  \im \int_k \frac{\bar{h}^{2}_{ab}(k+p)}{(k+p)^2}t^a\Sigma(k^2)\big[k^2-\Sigma^\dag(k^2)\Sigma(k^2)\big]^{-1}t^b \,,
\end{equation}
where for the flavor effective coupling we accept the heuristic Ansatz
\begin{eqnarray}\label{effective_charge}
\frac{\bar{h}^{2}_{ab}(q)}{q^2}
 & \stackrel{\mathrm{IR}}{\simeq} & \frac{h^{2}_*}{q^2}\Pi_{ac}(q)\big[1+\Pi(q)\big]_{cb}^{-1} \nonumber\\
 & \simeq & -h^{2}_*\frac{M_{ac}^{2}}{q^2}\big[q^2-M^{2}\big]_{cb}^{-1} \,,
\end{eqnarray}
where $h_*$ is a non-perturbative infrared fixed point of the flavor
gauge dynamics, $\Pi_{ab}(q)$ is the flavor gauge boson self-energy,
and $M_{ab}^{2}$ is the flavor gauge boson mass matrix.
(The rationale of the Ansatz is given in \cite{Benes:2011gi}.)

\subsection{Flavor symmetry breaking}

The flavor symmetry breaking and the fermion mass generation via
formation of the chirality changing self-energies are induced by
the strong flavor dynamics. Therefore it is essentially
non-perturbative phenomenon, hard to control. This fact is
condensed in the Schwinger--Dyson equation \eqref{SDE} and in our
impotence to solve it.

At least some qualitative understanding can be gained
if we treat the self-energies $\Sigma$, the flavor symmetry breaking
order parameters, as condensates formed by the pairing of the flavored
fermion chiral components.

In a regime of very high energies ($>\Lambda_\F$) the system is
fully symmetric, the flavor gauge bosons are massless and the
power of attraction, mediated by the massless flavor gauge bosons,
can be estimated by the Most Attractive Channel (MAC) method
\cite{Raby:1979my}.

The attractiveness of different pairing channels
\begin{equation}
\mathbf{r}_1\times \mathbf{r}_2\rightarrow \mathbf{r}_\mathrm{pair}
\end{equation}
is roughly measured by the quantity
\begin{equation}\label{AC}
\Delta C_2=C_2(\mathbf{r}_1)+C_2(\mathbf{r}_2)-C_2(\mathbf{r}_\mathrm{pair}) \,,
\end{equation}
where $C_2(\mathbf{r})$ is the quadratic Casimir invariant for the
representation $\mathbf{r}$, see Tab.~\ref{table} in the appendix
\ref{appB}.

Decreasing the energy scale, the attractiveness of different
pairing channels increases differently. Once the most attractive
channel produces the flavor symmetry breaking at the energy scale
$\Lambda_\F$, the MAC method looses its plausibility for the
remaining pairing channels since the flavor gauge bosons become
massive.

\subsubsection{Drawbacks of the minimal version}

The \emph{minimal} version analyzed in \cite{Benes:2011gi}, where
all fields are in triplets or anti-triplets, is approximately vector-like above
the huge scale $\Lambda_\F$ because there we can neglect QCD and
electroweak effects. The most attractive channel is
$\mathbf{3}\times\overline{\mathbf{3}}\rightarrow\mathbf{1}$ with
$\Delta C_2=8/3$. It causes several shortcomings of the minimal
version:


1) The most attractive channel is a flavor singlet, i.e., it
does not break the flavor symmetry. It suggests that the flavor
gauge dynamics should rather confine bellow $\Lambda_\F$.

2) Even if we assume that the QCD and electroweak dynamics are
sufficiently relevant at $\Lambda_\F$ to cure previous
shortcoming by inducing the necessary non-vector-like nature, it still
remains difficult to justify tininess of neutrino masses, simply,
because there is no natural reason for the see-saw pattern of neutrino mass matrix.

3) If at all, the breaking of the electroweak and the flavor
symmetry happens at once. The separation of the flavor scale
$\Lambda_\mathrm{F}$ and the electroweak symmetry breaking scale
$\Lambda_{\mathrm{EW}}\sim|\Sigma_u|$ is not obvious. Necessary
relation $\Lambda_\mathrm{F}\gg|\Sigma_u|$ has to be achieved by
critical scaling \cite{Miransky:1996pd,Braun:2010qs}.

\subsubsection{Advantages of the non-minimal version}

The \emph{non-minimal} version (63333) naturally and
straightforwardly leads to the complete flavor symmetry breaking
and cures the first two weak points immediately. On top of that
it provides the separation of flavor and electroweak symmetry
breaking. Requirement of the critical scaling, however, remains
unavoidable.

The attractive channels $(\mathrm{A.C.})$, governing different
parts of the neutrino self-energy written in the Nambu--Gorkov
formalism, are (compare with \eqref{NGselfenergy})
\begin{equation}
(\mathrm{A.C.})=\beginm{c|c|c}
    \mathbf{3}\times\mathbf{3}\rightarrow\overline{\mathbf{3}} & \ \ \ \ \overline{\mathbf{3}}\times\mathbf{3}\rightarrow\mathbf{1} \ \ \ \ &
                                                            \mathbf{6}\times\mathbf{3}\rightarrow\mathbf{8} \\ \hline
    \mathbf{3}\times\overline{\mathbf{3}}\rightarrow\mathbf{1} & \overline{\mathbf{3}}\times\overline{\mathbf{3}}\rightarrow\mathbf{3} &
                    \mathbf{6}\times\overline{\mathbf{3}}\rightarrow\mathbf{3} \vphantom{\begin{array}{c}\ \\ \ \end{array}}\\ \hline
    \mathbf{3}\times\mathbf{6}\rightarrow\mathbf{8} & \overline{\mathbf{3}}\times\mathbf{6}\rightarrow\mathbf{3} & \mathbf{6}\times\mathbf{6}\rightarrow\overline{\mathbf{6}} \endm \,.
\end{equation}
The measure \eqref{AC} of the attractiveness of the channels is
\begin{equation}
(\Delta C_2)=\beginm{c|c|c}
    4/3 & \ \ \ \ 8/3 \ \ \ \ & 5/3 \\ \hline
    8/3 & 4/3 & 10/3 \vphantom{\begin{array}{c}\ \\ \ \end{array}}\\ \hline
    5/3 & 10/3 & 10/3 \endm \,.
\end{equation}

It naturally follows that, decreasing the energy scale, the right-handed
neutrino pairing of Majorana type with $\Delta C_2=10/3$ happens first. This fact brings nice features:


1) It breaks the flavor symmetry providing no confinement.

2) It suggests the see-saw pattern of neutrino mass matrix.

3) It does not break the electroweak symmetry what is postponed to lower energies.

\subsubsection{Effective description of the flavor symmetry breaking}

We can quantify the anti-sextet and the four triplet pairings by,
so called, sterility condensates
\begin{subequations}
\begin{eqnarray}\label{condensates}
\bra{0}\frac{1}{4}\epsilon^{ACE}\epsilon^{BDF}\overline{(\xi_{\mathrm{R}}^{CD})^\C}\xi_{\mathrm{R}}^{EF}\ket{0} & \propto
& \Lambda_{\mathrm{F}}^2\bra{0}\Phi_{6}^{AB}\ket{0} \,,\hspace{0.5cm} \\
\bra{0}\overline{(\xi_{\mathrm{R}}^{AB})^\C}\zeta_{\mathrm{R}n}^{B}\ket{0} & \propto
& \Lambda_{\mathrm{F}}^2\bra{0}\Phi_{3}^{n,A}\ket{0} \,.\hspace{0.5cm}
\end{eqnarray}
\end{subequations}
where we have introduced auxiliary scalar fields $\Phi_6$ and
$\Phi^{n}_3$ of mass dimension one. The index $n=1,..,4$ is the
$\SU{4}_\mathrm{S}$ sterility index. The indices,
$A,B,C,..=1,..,3$, are the indices of the fundamental flavor
representation, and $\epsilon^{ABC}$ is the totally anti-symmetric
tensor. The auxiliary fields transform as an anti-sextet and a
triplet, respectively, under the flavor rotations ${\cal
U}=\e^{\im\alpha^a T^a}$
\begin{subequations}
\begin{eqnarray}
\Phi_{6}' & = & {\cal U}^{\dag\mathrm{T}}\Phi_{6}{\cal U}^{\dag} \,,\\
\Phi_{3}^{n}\vphantom{\Phi}' & = & {\cal U}\Phi_{3}^{n}\,.
\end{eqnarray}
\end{subequations}
These flavor transformation properties follow from the flavor transformation
properties of the elementary right-handed neutrino fields (for their definitions see \eqref{RHnu})
\begin{subequations}
\begin{eqnarray}
\xi_\mathrm{R}' & = & {\cal U}\xi_\mathrm{R}{\cal U}^{\mathrm{T}} \,,\\
{\zeta_\mathrm{R}^{n}}' & = & {\cal U}^*\zeta_\mathrm{R}^{n}\,,
\end{eqnarray}
\end{subequations}
and the fact that the totally anti-symmetric tensor $\epsilon^{ABC}$
is flavor invariant
\begin{equation}
{\cal U}^{AD}{\cal U}^{BE}{\cal U}^{CF}\epsilon^{DEF}=\epsilon^{ABC}\,.
\end{equation}
The quantum numbers $(\,L,\,S_3-S_6,\,S_3+S_6,\,\SU{4}_\mathrm{S}\,)$ of the scalar fields are
\begin{subequations}
\begin{eqnarray}
\Phi_{6}\ :\hspace{0.2cm} & & \left(\,2-2a,\,-2,\,+2,\,\mathbf{1}\,\right) \,, \\
\Phi_{3}^{n}\ :\hspace{0.2cm} & & \left(\,1-\frac{3}{2}a,\,-\frac{3}{4},\,+\frac{5}{4},\,\mathbf{4}\,\right) \,.
\end{eqnarray}
\end{subequations}

$\Phi_6$, and $\Phi_{3}^{n}$ are 18 complex scalar fields. They
can be expressed in terms of twice as many real scalar fields from which several are the Nambu--Goldstone fields
of broken flavor and sterility symmetries.

\begin{eqnarray}\label{Phi6}
\Phi_6(x) & = & \e^{-2\im\alpha(x)}\e^{+2\im\beta(x)} \times \\
     & & \hspace{-1cm} \times\ \e^{-\im\theta^a(x) T^{a\mathrm{T}}}\beginm{ccc}\Delta_1(x) & 0 & 0 \\
                                                                0 & \Delta_2(x) & 0 \\
                                                                0 & 0 & \Delta_3(x) \endm\e^{-\im\theta^a(x) T^{a}} \,, \nonumber
\end{eqnarray}
\begin{eqnarray}\label{Phi3}
\beginm{c}\Phi^{1}_3(x)^\T \\ \Phi^{2}_3(x)^\T \\ \Phi^{3}_3(x)^\T \\ \Phi^{4}_3(x)^\T \endm & = & \e^{-\frac{3}{4}\im\alpha(x)}\e^{+\frac{5}{4}\im\beta(x)}\e^{\im\gamma^i(x)s^{i}}\times \\
          & & \hspace{-2cm}\times\ \e^{\im\theta^a(x) T^{a}}\beginm{ccccc}
                        \big(\hspace{-5pt} & 0 & 0 & 0 & \hspace{-5pt}\big) \\
                        \big(\hspace{-5pt} & 0 & 0 & \delta_2(x) & \hspace{-5pt}\big) \\
                        \big(\hspace{-5pt} & 0 & \delta_3(x) & \delta_1(x) & \hspace{-5pt}\big) \\
                        \big(\hspace{-5pt} & \delta_4(x) & \varepsilon_5(x) & \varepsilon_6(x)   & \hspace{-5pt}\big)
                    \endm \,. \nonumber
\end{eqnarray}

The 25 Nambu--Goldstone bosons are (for majorons see section \ref{secV}):
\begin{itemize}
\item 8 of $\theta^a(x)$ corresponding to broken $\SU{3}_\mathrm{F}$: \\
    longitudinal components of flavor gauge bosons $C^{\mu}_L$
\item 15 of $\gamma^i(x)$ corresponding to broken $\SU{4}_\mathrm{S}$: \\
    non-Abelian light majorons $\eta^i$
\item 1 of $\alpha(x)$ corresponding to broken $S_3-S_6$: \\
    Abelian light majoron $\eta^0$
\item 1 of $\beta(x)$ corresponding to broken $S_3+S_6$: \\
    super-heavy majoron $H$
\end{itemize}
Further there are 7 real and 2 complex scalars and in general they all can develop 9 CP-preserving vacuum expectation values $\phi_A$ and $\varphi_k$ and 2 CP-violating phases $\varsigma_k$.
\begin{itemize}
\item 3 real components of sextet field \\ $\Delta_A(x)\rightarrow\Delta_A(x)+\phi_A$, $A=1,2,3$
\item 4 real anti-triplet fields \\
    $\delta_k(x)\rightarrow\delta_k(x)+\varphi_k$, $k=1,2,3,4$
\item 2 complex anti-triplet fields \\ $\varepsilon_k(x)\rightarrow\varepsilon_k(x)+\varphi_k\e^{\im\varsigma_k}$, $k=5,6$
\end{itemize}

A general form of the condensate $\bra{0}\Phi_{6}\ket{0}$ is
\begin{equation}\label{cond6}
\bra{0}\Phi_{6}\ket{0}=\beginm{ccc}   \phi_1 & 0 & 0 \\
                                      0 & \phi_2 & 0 \\
                                      0 & 0 & \phi_3 \endm
\end{equation}
as follows from \eqref{Phi6}. A general form of the condensates
$\bra{0}\Phi_{3}^{n}\ket{0}$ follows from \eqref{Phi3}. In general they are complex and have
nontrivial mutual angle and also non-trivial angle with respect
to the $\bra{0}\Phi_{6}\ket{0}$.

Not only for the sake of concreteness we choose here a special form of the triplet condensates
\begin{subequations}\label{cond3}
\begin{eqnarray}
\bra{0}\Phi_{3}^{n=1,2,3}\ket{0} & = & \ \ 0 \,, \\
\bra{0}\Phi_{3}^{n=4}\ket{0}\ \  & = & \beginm{ccc}\varphi_4 & \varphi_5 & \varphi_6 \endm \,.
\end{eqnarray}
\end{subequations}
The main reason for this choice is that it leaves the $\SU{3}_\mathrm{S}$ sterility subgroup unbroken what is necessary to protect the seesaw mechanism (see Sect.~\ref{neutrino_pheno}). Without the special form the general condensates would break the sterility symmetry completely.

The condensates break the flavor symmetry completely while the
electroweak symmetry breaking is postponed to the lower energies
where the pairing of the electroweakly charged fermions occurs.
The sextet sterility condensation is very similar to the sextet
color superconductivity \cite{Brauner:2003pj}.

\subsubsection{Masses from the sterility condensation}

The sterility condensation produces masses of all flavor gauge
bosons. The masses can be estimated from the lowest order gauge
invariant kinetic terms of the effective Lagrangian for the
effective scalar fields
\begin{equation}\label{L_M_gauge}
{\cal L}_{\mathrm{M_{gauge}}}=(D^\mu\Phi_{3}^{n})^\dag D_\mu\Phi_{3}^{n}+\Tr(D^\mu\Phi_{6})^\dag D_\mu\Phi_{6} \,,
\end{equation}
where
\begin{subequations}
\begin{eqnarray}
D_\mu\Phi_{6} & = & \partial_\mu\Phi_{6}+\im h C^{a}_\mu(T^{a\mathrm{T}}\Phi_{6}+\Phi_{6}T^{a}) \,,\\
D_\mu\Phi_{3}^{n} & = & (\partial_\mu-\im h C^{a}_\mu T^a)\Phi_{3}^{n} \,.
\end{eqnarray}
\end{subequations}

In the effective Lagrangian ${\cal L}_{\mathrm{M_{gauge}}}$
\eqref{L_M_gauge} we substitute the effective scalar fields for
their vacuum expectation value $\Phi\rightarrow\bra{0}\Phi\ket{0}$,
and we get the mass matrix for the gauge bosons
\begin{equation}
M_{\mathrm{gauge}}^2=M_{6}^2+M_{3}^2 \,,
\end{equation}
where the mass matrices $M_{6}^2$ and $M_{3}^2$ with the specific
form of the condensates \eqref{cond6} and \eqref{cond3}
are in the Appendix \eqref{Mgauge}.

The sterility condensation produces also Majorana masses for
right-handed neutrinos. The masses can be estimated from
Yukawa terms of the effective Lagrangian for the effective
scalar fields
\begin{eqnarray}\label{L_M_sterile}
{\cal L}_{\mathrm{M_{R}}} & = &
\hphantom{+\ }y_{36}\,\overline{(\zeta_{\mathrm{R}}^{n})^\C}\xi_{\mathrm{R}}\Phi_{3}^{n*} \\
& & +\ y_{6}\,\epsilon^{ACE}\epsilon^{BDF}\overline{(\xi_{\mathrm{R}}^{AB})^\C}\xi_{\mathrm{R}}^{CD}(\Phi_{6}^{EF})^\dag  \nonumber \\
& & +\ \hc \,,\nonumber
\end{eqnarray}
where the effective Yukawa coupling constants
\begin{equation}
y_{36}=\frac{4}{9}h^2 \nonumber \,,\
y_{6}=\frac{4}{9}h^2
\end{equation}
are obtained from the effective four-neutrino interaction
$\sim(\bar{n}\gamma_\mu t^an)(\bar{n}\gamma^\mu t^an)$.

In the effective Lagrangian ${\cal L}_{\mathrm{M_{R}}}$
\eqref{L_M_sterile} we can substitute the scalars for the condensates and get the Majorana
mass matrix for the sterile neutrinos\footnote{Here the condensates should be rewritten in the $\nu_{\mathrm{R}}$-formalism \eqref{NGmultiplet}, not in the matrix $\xi_\mathrm{R}$-formalism. }
\begin{equation}\label{Msterile}
M_\mathrm{R}=\frac{4}{9} h^2\beginm{c|c}
    0 & \langle\Phi^{n}_3\rangle \vphantom{\begin{array}{c}\ \\ \ \end{array}}\\ \hline
    \ \langle\Phi^{m}_3\rangle^\mathrm{T} & \langle\Phi_6\rangle \\ \endm \,.
\end{equation}
The mass matrix $M_\mathrm{R}$ has generically at least six zero
eigenvalues. With the special choice of sterility condensates
\eqref{cond6} and \eqref{cond3}, there are nine zero eigenvalues.

\subsection{Neutrino phenomenology}\label{neutrino_pheno}

The neutrino masses are given as roots of the equation \eqref{detSigma}
where the momentum dependence of $\Sigma(p^2)$ makes the calculation difficult.
For qualitative purpose it is sufficient to substitute the self-energy by
a \emph{constant} $N\times N$ symmetric mass matrix $M$, where $N=21$ for our case.

The mass spectrum can be found as eigenvalues $m_1,..,m_N$ of $M$,\footnote{If the mass matrix $M$ is complex then we have to find eigenvalues of $M^\dag M$ to determine the mass spectrum.}
\begin{subequations}
\begin{eqnarray}
\hspace{-1.5cm} & & \overline{\nu^\C} M \nu  =  \overline{\nu^\C} U^\T \beginm{ccc}  \vspace{-0.2cm} m_1 & & \\ \hspace{-0.3cm}\vspace{-0.2cm} & \ddots & \\ \hspace{-0.3cm} & & m_N\endm U\nu \\
\hspace{-0.1cm} & & \ \ \stackrel{\mathrm{e.g.}}{=}
\left(\overline{(\nu_{1}')^\C} \ldots\ \overline{(\nu_{N}')^\C}\right)
\hspace{-0.1cm}\beginm{ccccc} \vspace{-0.2cm} 0 & & & & \\ \hspace{-0.4cm}\vspace{-0.2cm} & m & & & \\ \hspace{-0.5cm}\vspace{-0.2cm} & & m & & \\ \hspace{-0.7cm}\vspace{-0.2cm} & & & \ddots & \\ \hspace{-0.7cm} & & & & m' \endm\hspace{-0.2cm} \beginm{c}\nu_{1}'\\ \vdots\\\nu_{N}'\endm  \,, \nonumber\\
& &  \label{egMassMatrix}
\end{eqnarray}
\end{subequations}
where $U$ is the diagonalizing unitary transformation matrix.

Three types of mass eigenstates can arise: (i) In the most general scenario, when
no selection rule is in work, all eigenvalues come out nonzero and different. In our
case, they correspond to 21 massive Majorana neutrinos.
(ii) The zero eigenvalues correspond to massless Weyl neutrinos. (iii) It can happen
that pairs of degenerate eigenvalues appear (see e.g. in \eqref{egMassMatrix}). Each
pair then corresponds to a massive Dirac neutrino with its chiral components given as,
e.g.,
\begin{subequations}
\begin{eqnarray}
\nu_\mathrm{L} & = & \nu_{2}'+\im\nu_{3}' \,,\\
\nu_\mathrm{R} & = & (\nu_{2}')^\C-\im(\nu_{3}')^\C \,.
\end{eqnarray}
\end{subequations}
The presence of the pair degeneracy signals the
$\OO{2}\sim\U{1}$ symmetry of the mass matrix, a subgroup of either flavor, or
sterility, or both. The symmetry corresponds to a quantum number carried by the Dirac
neutrino.

In the minimal version (333), the left- and right-handed Majorana, and Dirac entries of the
neutrino mass matrix arise from the pairing channels of the same flavor structure
$\mathbf{3}\times\mathbf{3}$ or $\overline{\mathbf{3}}\times\overline{\mathbf{3}}$,
thus of the same strength of attraction. That does not indicate
the see-saw pattern of the neutrino mass matrix at all. The reason for the tiny masses
of electroweak neutrinos has to be fully left on a huge amplification effects \cite{Benes:2011gi}.
It is then natural to expect that the remaining nine sterile neutrinos turn out to be
of small mass as well. In the same time, the dynamics should be also responsible for sufficient
suppression of the right-handed admixtures within the electroweak neutrinos, for what it is
difficult to find some natural reason.

On contrary, the non-minimal version (63333) naturally leads to the dynamically generated
see-saw pattern of the neutrino mass matrix. The see-saw pattern is useful
not only for explanation of tiny masses of the electroweak neutrinos, but
also for suppression of their oscillations into the sterile neutrinos.

The key point is the presence of the sextet right-handed neutrinos. Their privileged
role makes the situation clearer, separates the study of the right-handed neutrinos
from other fermions, and allows us to switch into the approximative description by
condensates.

Within the (63333) version we have demonstrated so far the massiveness only of the
right-handed neutrinos. But of course we expect that ultimately at lower
energy scale all elements of the full neutrino mass matrix given by \eqref{NGselfenergy}
become non-vanishing and all mass eigenstates become massive. The sole fact that
there is an odd number of neutrino degrees of freedom indicates that at least
one neutrino must be of Majorana type.

By the construction above we want to show that the right-handed Majorana
elements dominate the whole neutrino mass matrix. This is exactly what
is needed for the see-saw mechanism to work. Due
to the strength of the sextet neutrino condensation the see-saw pattern
occurs dynamically and naturally.

Nevertheless, the system with the general sterility condensation scheme is not
directly able to accommodate all three light electroweak neutrinos.
The see-saw mechanism is triggered by switching on the Dirac elements of the neutrino mass matrix.
It seems to be natural to switch on the next-to-most attractive channel, $D^{\overline{\mathbf{3}}\times\mathbf{3}}_n$
\eqref{NGselfenergy}. They arise dynamically from the effective four-fermion interactions
\begin{equation}
\left[h^2/M^{2}_\mathrm{gauge}\right]_{ab}(\bar\nu_\mathrm{L}\gamma_\mu T^{a*}\nu_\mathrm{L})(\bar\zeta_{\mathrm{R}}^{n}\gamma^\mu T^{b*}\zeta_{\mathrm{R}}^{n})
\end{equation}
after appropriate Fiertz rearrangement.
Because it has analogous flavor structure as $u$-quark mass matrix,
$\overline{\mathbf{3}}\times\mathbf{3}$, we assume it to be of the same order of magnitude,
$D^{\overline{\mathbf{3}}\times\mathbf{3}}_n\sim m_t$. This should provide the see-saw
masses of the electroweak neutrinos, $m_{\nu_\mathrm{EW}}\sim m_{t}^2/\Lambda_\F$,
and of the sterile neutrinos, $m_{\nu_\mathrm{sterile}}\sim \Lambda_\F$. To reproduce the
light neutrino mass $m_{\nu_\mathrm{EW}}\sim10^{-1}\,\mathrm{eV}$ while
$m_{t}\sim10^{2}\,\mathrm{GeV}$, the flavor scale should be at least $\Lambda_\F\sim10^{14}\,\mathrm{GeV}$.

In our system with general $D^{\overline{\mathbf{3}}\times\mathbf{3}}_n\sim m_t$
and $R^{\overline{\mathbf{3}}\times\mathbf{6}}_1\sim\Lambda_\F$, the seesaw mechanism does not work.
It is caused by the presence of six zero eigenvalues of general $M_\mathrm{R}$ \eqref{Msterile}.
Instead of combining with super-heavy modes producing the seesaw spectrum, the three left-handed neutrino modes combine with three of the right-handed neutrino zero-modes to produce three pairs of quasi-degenerate modes of mass ($\sim m_t$). Those
six modes in fact appear as three, too heavy pseudo-Dirac electroweak neutrinos in flagrant
contradiction with observations.

There is way out of this trouble, if a subgroup of the sterility symmetry, which is able to prohibit
the mixing of the left-handed neutrinos with the right-handed zero-modes, remains unbroken.
The see-saw mechanism then acts only on the left-handed and super-heavy right-handed neutrino
modes. All the way down to lower energy scale, the residual symmetry is unbroken and keeps
the right-handed zero modes massless and decoupled from the massive neutrinos. We can say
that we need the residual symmetry to protect the see-saw mechanism.

The necessary residual sterility symmetry can be achieved by imposing the special form \eqref{cond3}
of the triplet condensation which is equivalent to dynamically natural relations that $\bra{0}\Phi_{3}^{n}\ket{0}=\bra{0}\Phi_{3}^{n'}\ket{0}$ (or generally $R^{\overline{\mathbf{3}}\times\mathbf{6}}_n=R^{\overline{\mathbf{3}}\times\mathbf{6}}_{n'}$),
and $D^{\overline{\mathbf{3}}\times\mathbf{3}}_n=D^{\overline{\mathbf{3}}\times\mathbf{3}}_{n'}$, see
\eqref{NGselfenergy}. The seesaw-mechanism-protecting residual sterility symmetry is then $\SU{3}_\mathrm{S}\subset\SU{4}_\mathrm{S}$ generated by $S_i$, with $i=1,..,8$.


\section{Majorons}\label{secV}
Experienced by QCD we expect that the strong flavor dynamics leads to
rich bound state spectrum. Their complete description and classification
is nevertheless an infeasible task. The only bound states we can
be sure to exist are the Nambu--Goldstone bosons of the spontaneously
broken global symmetries. In this section we concentrate on the
classification of bound states that arise from the formation of the
neutrino self-energy of general form, spontaneously breaking the lepton number and
the sterility symmetry $G_\mathrm{S}$, see section \ref{nuSymmetries}.

We leave the line of the previous section where, by phenomenological preferences, we have been lead
to the particular pattern of the neutrino self-energy. We assume general pattern providing maximal chiral
symmetry breaking occurring at one energy scale $\Lambda_\F$. The sterility
symmetry together with the lepton number is broken completely along with the
flavor symmetry breaking and rich spectrum of Nambu--Goldstone bosons, so called
majorons, appears. For free the model provides excellent scalar candidates
for the dark matter \cite{Berezinsky:1993fm}. Later in this section we
describe and classify them.

The majoron corresponding to the anomalous symmetry is not true
Nambu--Goldstone boson. It rather acquires huge mass ($\sim\Lambda_\mathrm{F}$),
analogously to the case of the $\eta'$ in QCD. This majoron is called
the \emph{heavy sterile} majoron $H$.

The majorons corresponding to the anomaly free part of the sterility
symmetry are called the \emph{light sterile} majorons $\eta$.

The spontaneously broken anomaly free lepton number $L$ gives rise to the
\emph{standard} majoron $J$ \cite{Chikashige:1980ui,Schechter:1981cv}.
It is always present in all versions of the model.

The majorons $\eta$ and $J$ are the true Nambu--Goldstone bosons.
They nevertheless do not present a phenomenological problem
in the form of new long range force. The argument is simple:
The Nambu--Goldstone bosons mediate spin-dependent tensor force
among fermions which vanishes with cube of distance \cite{Gelmini:1982zz}.

What more, the majorons can eventually acquire mass by gravitational effects of the
order of, say, few keV \cite{Coleman:1988tj,Giddings:1988cx,Akhmedov:1992hi}.
That would of course drastically shorten the force range.
In the formulation of the issue we omit these effects and treat the majorons $\eta$ and $J$
as massless. During the phenomenological analysis, nevertheless, we keep this
possibility open and call them collectively as \emph{light} majorons.


\subsection{Light majorons}

All versions of the model predict the existence of the standard majoron $J$ from the spontaneously
broken lepton number $L$ \eqref{Lepton_number}. The standard majoron couples to the lepton number current
\begin{equation}
\bra{0}\mathcal{J}^{\mu}_{\mathrm{L}}(0)\ket{J(q)}=\im q^\mu F_J \\
\end{equation}
with the strength of the standard majoron decay constant $F_J$.
The anomaly free lepton number is spontaneously broken by the formation
of all Dirac, $\Sigma_\mathrm{D}$, left-handed Majorana, $\Sigma_\mathrm{L}$,
and right-handed Majorana, $\Sigma_\mathrm{R}$, components of the neutrino
self-energy. Therefore the standard majoron is created from vacuum by a
linear combination of interpolating operators
\begin{equation}
J\sim\left(\overline{\nu_\mathrm{R}}\nu_\mathrm{L}+\overline{\nu_\mathrm{L}}\nu_\mathrm{R}\right),\ \  \overline{\nu^{\C}_\mathrm{L}}\nu_\mathrm{L},\ \  \overline{\nu^{\C}_\mathrm{R}}\nu_\mathrm{R} \,.
\end{equation}

In the version (63333) sixteen light sterile majorons $\eta_0$ and $\eta_i$, $i=1,..,15$, as the
Nambu--Goldstone bosons couple to the sterility currents
\begin{subequations}
\begin{eqnarray}
\bra{0}\mathcal{J}^{\mu}_{\mathrm{S}_3-\mathrm{S}_6}(0)\ket{\eta_0(q)} & = & \im q^\mu F_\mathrm{\eta} \,;\\
\bra{0}\mathcal{J}^{\mu}_{\mathrm{S},i}(0)\ket{\eta_j(q)} & = & \im q^\mu F_\mathrm{\eta}\delta_{ij}
\end{eqnarray}
\end{subequations}
with the strength of the sterile majoron decay constant $F_\mathrm{\eta}$.
The sterility symmetry is spontaneously broken by the formation of both
Dirac, $\Sigma_\mathrm{D}$, and right Majorana, $\Sigma_\mathrm{R}$, components of
the neutrino self-energy. Therefore the sterile majorons are created from vacuum by a linear
combination of interpolating operators
\begin{subequations}\label{Inter_fields}
\begin{eqnarray}
\eta_i & \sim & \left(\overline{\nu_\mathrm{R}}S_i\nu_\mathrm{L}+\overline{\nu_\mathrm{L}}S_i\nu_\mathrm{R}\right),\ \ \overline{\nu^{\C}_\mathrm{R}}S_i\nu_\mathrm{R} \,;\\
\eta_0 & \sim & \left(\overline{\nu_\mathrm{R}}\nu_\mathrm{L}+\overline{\nu_\mathrm{L}}\nu_\mathrm{R}\right),\ \ \overline{\nu^{\C}_\mathrm{R}}\nu_\mathrm{R} \,.\label{Inter_fields_C}
\end{eqnarray}
\end{subequations}

In the following we will use the common notation for the generators relevant
for the light majorons, $X_\alpha$, $\alpha=0,1,..,16$. It denotes the vector of the lepton number
and sterility generators
\begin{subequations}\label{Z_generators}
\begin{eqnarray}
S_3-S_6: & & X_0=s_0 \,; \\
\SU{4}_\mathrm{S}: & & X_i=s_i \,,\ \mathrm{where}\ i=1,..,15\,; \\
L: & & X_{16}=l\,.
\end{eqnarray}
\end{subequations}
where the lepton number generator in the Nambu--Gorkov formalism $l$ is
introduced in \eqref{NGleptonNumberGenerator}, and the sterility symmetry
generators in the Nambu--Gorkov formalism $s_\alpha$ are introduced in
\eqref{NGsterileGenerator}.

Not counting their mutual interactions, the majorons interact mainly
with neutrinos. Such interactions can be described generally by effective
Yukawa majoron-neutrino term
\begin{subequations}\label{EffYukawa}
\begin{eqnarray}
{\cal L}_{\mathrm{eff,}\eta\nu\nu} & = & y_{\eta\nu\nu}\,(\bar n X_0 n)\,\eta_0 + y_{\eta\nu\nu}'\,(\bar n X_i n)\,\eta_i \,,\\
{\cal L}_{\mathrm{eff,}J\nu\nu} & = & y_{J\nu\nu}\,(\bar n X_{16} n)\,J \,.
\end{eqnarray}
\end{subequations}
At that level the effective Yukawa coupling constants $y_{\eta\nu\nu}$, $y_{\eta\nu\nu}'$ and $y_{J\nu\nu}$
are mere parameters. Nevertheless the majoron-neutrino coupling strength can be related to
more fundamental quantities of the model, like to the flavor symmetry breaking neutrino
self-energy $\mathbf{\Sigma}(p)$.

For that purpose we follow standard procedure \cite{Jackiw:1973tr,Cornwall:1973ts}
to insist on the fulfilment of the Ward--Takahashi identity for a proper vertex
$\Gamma^{\mu}_\alpha(p+q,p)$ corresponding to the Green functions
$G^{\mu}_0(x,y,z)\equiv\bra{0}T\mathcal{J}^{\mu}_{\mathrm{S}_3-\mathrm{S}_6}(x)n(y)\bar{n}(z)\ket{0}$,
$G^{\mu}_i(x,y,z)\equiv\bra{0}T\mathcal{J}^{\mu}_{\mathrm{S},i}(x)n(y)\bar{n}(z)\ket{0}$ and
$G^{\mu}_{16}(x,y,z)\equiv\bra{0}T\mathcal{J}^{\mu}_{L}(x)n(y)\bar{n}(z)\ket{0}$
of the lepton number and sterility currents coupled to the neutrino fields. The
Ward--Takahashi identity reads
\begin{equation}\label{WTI}
q_\mu\Gamma^{\mu}_\alpha(p+q,p)=S^{-1}(p+q)X_\alpha-\gamma_0 X_\alpha\gamma_0
S^{-1}(p) \,.
\end{equation}
When the dynamics develops symmetry breaking neutrino self-energy,
the Ward--Takahashi identity \eqref{WTI} does not vanish for $q\rightarrow0$.
It determines uniquely only the leading ${\cal O}(q^{-1})$ part $\Gamma^{\mu}_{\alpha,\mathrm{lead.}}$
of the proper vertex $\Gamma^{\mu}_\alpha$
\begin{equation}\label{Gamma_pole}
\Gamma^{\mu}_\alpha(p+q,p)=\Gamma^{\mu}_{\alpha,\mathrm{lead.}}(p+q,p)+{\cal O}(q^{0}) \,,
\end{equation}
where
\begin{equation}
\Gamma^{\mu}_{\alpha,\mathrm{lead.}}(p+q,p)=-\frac{q^\mu}{q^2}\left(\mathbf{\Sigma}(p)X_\alpha-\gamma_0 X_\alpha\gamma_0\mathbf{\Sigma}(p)\right) \,.
\end{equation}
Physically, we interpret the pole in terms of the exchange of the massless Nambu--Goldstone boson,
i.e., majoron. Following this interpretation we pick up the Nambu--Goldstone part of
the proper vertex
\begin{equation}\label{Gamma_NG}
\Gamma^{\mu}_{\alpha}(p+q,p)=\Gamma^{\mu}_{\alpha,{\mathrm{NG}}}(p+q,p)+\ldots
\end{equation}
and approximate it by a `one-loop' expression
\begin{eqnarray}
\Gamma^{\mu}_{\alpha,{\mathrm{NG}}}(p+q,p) & \approx & \begin{array}{c}\includegraphics[width=0.3\textwidth]{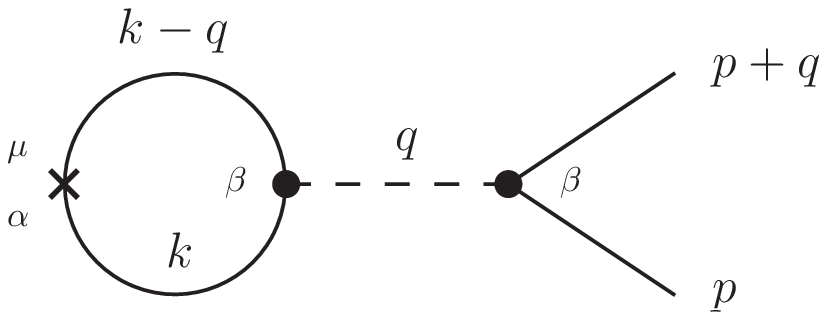}\end{array}  \nonumber\\
                   & = & -\frac{q^\mu}{q^2}I_{\alpha\beta}(q^2)P_\beta(p+q,p) \,,\label{Gamma_NG:loop}
\end{eqnarray}
where the massless majoron propagator $\tfrac{\delta_{\beta\gamma}}{q^2}$ connects the neutrino
loop $q^\mu I_{\alpha\beta}(q^2)$ and the majoron-neutrino vertex $P_\gamma(p+q,p)$ both regular for $q=0$.

Comparing the two expressions \eqref{Gamma_pole} and \eqref{Gamma_NG} for $q\rightarrow0$ we get
the expression for the majoron-neutrino vertex for $q=0$
\begin{equation}\label{P_J_nu}
P_\alpha(p,p)=I^{-1}_{\alpha\beta}(0)\big[\mathbf{\Sigma}(p)X_\beta-\gamma_0 X_\beta\gamma_0\mathbf{\Sigma}(p)\big] \,.
\end{equation}
The loop function $I_{\alpha\beta}(0)$ from the diagram \eqref{Gamma_NG:loop} is
\begin{equation}\label{I_def}
I_{\alpha\beta}(0)  =
\im\lim_{q\rightarrow0}\int_k\Tr\left(\frac{\slashed{q}}{q^2}X_\alpha S(k-q)P_\beta(k-q,k)S(k)\right) \,.
\end{equation}
For the sake of simplicity we write explicit formula for the loop function $I_{\alpha\beta}(0)$
only within the approximation of constant self-energies, $\mathbf{\Sigma}(p)\rightarrow \mathbf{M}\equiv\mathbf{\Sigma}(0)$. The
approximated $I_{\alpha\beta}(0)$ then if plugged into \eqref{P_J_nu} gives us an upper estimate
of magnitude of the majoron-neutrino vertex $P_\alpha(p,p)$.

Due to the limit in \eqref{I_def},
we need to expand the $q$-dependent quantities up to ${\cal O}(q^1)$ order. The expansion
of the neutrino propagator is
\begin{equation}
\tilde{S}(k-q)=\tilde{S}(k)+\tilde{S}(k)\slashed q \tilde{S}(k)+{\cal O}(q^2)
\end{equation}
and we \emph{assume} that the expansion for the majoron-neutrino vertex is
\begin{equation}
\tilde{P}_\beta(k-q,k)=\tilde{P}_\beta(k,k)+{\cal O}(q^2) \,,
\end{equation}
where the tilde means the constant self-energy approximation of the quantity.

The loop function $I_{\alpha\beta}(0)$ necessary for the majoron-neutrino vertex
\eqref{P_J_nu} is given by relation
\begin{eqnarray}
\left[I(0)I^{\T}(0)\right]_{\alpha\beta} & = &
\im\lim_{q\rightarrow0}\int_k\Tr\left(\frac{\slashed{q}}{q^2}X_\alpha \tilde{S}(k)\slashed q \tilde{S}(k)\times\right. \\
& & \hspace{1cm} \left.\vphantom{\frac{\slashed{q}}{q^2}}
\times\big[\mathbf{M} X_\beta-\gamma_0
X_\beta\gamma_0 \mathbf{M}\big]\tilde{S}(k)\right) \,. \nonumber
\end{eqnarray}

\subsection{Heavy sterile majoron}
The sterile majoron $H$ couples to the anomalous current of Abelian sterility
symmetry $\U{1}_{\mathrm{S}_3+\mathrm{S}_6}$
\begin{equation}
\bra{0}\mathcal{J}_{\mathrm{S}_3+\mathrm{S}_6}^\mu(0)\ket{H(q)}=\im q^\mu F_\mathrm{S} \,.
\end{equation}
The heavy sterile majoron is created from vacuum by neutrino interpolating fields \eqref{Inter_fields_C}, and
additionally, by flavor gauge boson component which is a topologically
nontrivial field configuration
\begin{equation}
H\sim F_{\mu\nu a}\tilde{F}^{\mu\nu}_a \,.
\end{equation}

The majoron $H$ acquires huge mass due to the strong flavor axial anomaly
\eqref{anomaly}. The value of its mass can be estimated according to the
$\eta'$ mass analysis in QCD \cite{Witten:1979vv,Veneziano:1979ec,Shore:1998dm} as
\begin{equation}
m_{H}^2\sim\frac{\chi(0)}{F^{2}_\mathrm{S}}\sim\Lambda_{\mathrm{F}}^2 \,,
\end{equation}
where the flavor topological susceptibility is
estimated as $\chi(0)\sim\Lambda^{4}_\mathrm{F}$,
and the decay constant as $F_{\mathrm{S}}\sim\Lambda_\mathrm{F}$.

The anomalous coupling of $H$ to the flavor gauge bosons is given as
\begin{equation}\label{axionEWinteraction}
{\cal L}_{HCC}= \frac{h^2}{32\pi^2}\frac{H}{F_{\mathrm{S}}}F_{\mu\nu
a}\tilde{F}^{\mu\nu}_a \,.
\end{equation}

The effective coupling of the heavy majoron with the neutrinos is
\begin{equation}\label{L_Hnn}
{\cal L}_{Hnn}\sim \frac{m_n}{F_\mathrm{S}}H\bar{n}\gamma_5n \,.
\end{equation}
Because the interaction is proportional to the neutrino mass, the only
significant interaction is with the heavy sterile neutrinos, the
heavy sterile majoron is fairly invisible.

\subsection{Majoron phenomenology}

\subsubsection{Light majorons:}

Suppose that light majorons have mass of the order of few keV due to the
gravitational effects. Then they are suitable candidates for warm dark matter
\cite{Lattanzi:2008zz}. Important characteristic is their decay width. They
can decay only to $N_\mathrm{light}$ sufficiently light neutrinos with mass
$m_\mathrm{light}<M_J/2$, i.e., at least to the three electroweak neutrinos.

In the following we omit the differences among the light majorons and estimate
the decay width only for standard majoron $J$ of mass $M_J$. The matrix element
${\cal M}$ for the decay is simply governed by the effective majoron-neutrino interaction
\eqref{EffYukawa}:
\begin{equation}
\im{\cal M}=\begin{array}{c}\includegraphics[width=0.2\textwidth]{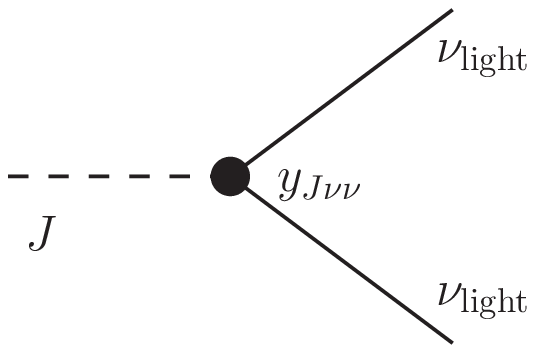}\end{array} \,.
\end{equation}
The decay width $\Gamma$ is given by
\begin{equation}\label{DW}
\Gamma(J\rightarrow nn)=\frac{N_\mathrm{light}}{8\pi}y_{J\nu\nu}^2 M_J\left(1-\frac{4m_\mathrm{light}^2}{M_{J}^2}\right)^{3/2} \,,
\end{equation}
where the effective Yukawa coupling is given by the majoron-neutrino vertex $P(p,p)$
\begin{equation}
y_{J\nu\nu}\sim P(p,p)\approx\frac{m_\mathrm{light}}{\sum_\nu m_\nu} \,.
\end{equation}
We have estimated the neutrino loop $I(0)$ \eqref{I_def} by a sum of all neutrino
mass eigenvalues $m_\nu$, $I(0)\approx \sum_\nu m_\nu$.

Now, in the version (333), we could expect that masses of all neutrino eigenstates turn
out to be of the same order, thus of the order of the electroweak neutrino mass. That is why
we can estimate the effective Yukawa coupling as $y_{J\nu\nu}^{(333)}\approx 10^{-1}$ and
neglect ratio $\tfrac{m_\mathrm{light}}{M_{J}}$. For the decay width we get an estimate
\begin{equation}
\Gamma^{(333)}(J\rightarrow nn)\approx10^{-3}M_J \,.
\end{equation}

On the other hand, in the version (63333) where the see-saw mechanism is in work, we could
expect that only the $N_\mathrm{light}=3$ electroweak neutrinos are very light, $m_\mathrm{light}\ll M_J$.
Then the decay width \eqref{DW} becomes
\begin{equation}
\Gamma^{(63333)}(J\rightarrow nn)\sim\frac{N_\mathrm{light}}{N_\mathrm{heavy}^2}\frac{m_\mathrm{light}^2}{\Lambda_{\F}^2}\frac{M_J}{8\pi}\approx10^{-50}M_J \,,
\end{equation}
where the sum of neutrino mass eigenvalues is dominated by $N_\mathrm{heavy}$ super-heavy neutrinos
of mass $\sim\Lambda_\F\approx10^{14}\,\mathrm{GeV}$.

That makes a qualitative difference between the two versions of the model. While in the version (333)
the light majorons are short-lived, in the version (63333) the light majorons are practically stable.
From this point of view the version (333) resembles more the triplet majoron models \cite{Gelmini:1980re},
while the version (63333) resembles the singlet majoron models \cite{Chikashige:1980ui}.

\subsubsection{Heavy sterile majoron:}

The coupling of the heavy sterile majoron to the flavor anomaly has important
consequences for the $CP$ properties of the flavor model. There is no reason
why there should not be the $\theta$-term of flavor gauge dynamics in the
effective Lagrangian. The $\theta$-parameter shifted by phase that makes the
neutrino masses real is eliminated by the Peccei--Quinn mechanism \cite{Peccei:1977hh,Peccei:1977ur},
where the heavy sterile majoron plays a role of the composite axion.

The heavy sterile majoron could decay to the heavy flavor gluons due to the
direct interaction \eqref{axionEWinteraction} induced by the flavor anomaly.
The decay would be kinematically allowed if the heavy sterile majoron is
heavier than twice the mass of $N_C\leq8$ lighter flavor gauge bosons, $M_H<2M_C$.
For the sake of rough estimate of the decay width, we omit the non-Abelian
character of the flavor gauge bosons and also the differences of their masses
using a common mass $M_C$. The matrix element ${\cal M}$ is given by the vertex
\eqref{axionEWinteraction}
\begin{eqnarray}
\im{\cal M} & = & \begin{array}{c}\includegraphics[width=0.2\textwidth]{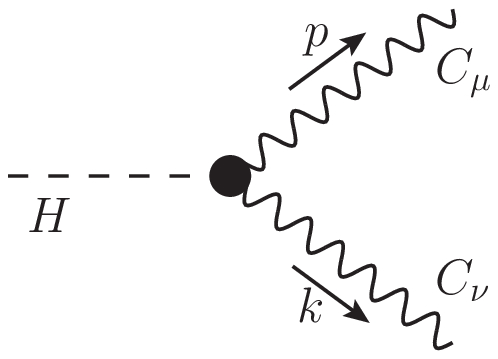}\end{array} \nonumber\\
    & = & \frac{h^2}{32\pi^2F_\mathrm{S}}\varepsilon_{\mu}^*(p)\varepsilon_{\nu}^*(k)\epsilon^{\mu\nu\alpha\beta}p_\alpha k_\beta \,,
\end{eqnarray}
where $\varepsilon_{\mu}$ is the polarization vector of the flavor gauge boson $C_\mu$.
The decay width then follows
\begin{equation}\label{DW_HM_CC}
\Gamma(H\rightarrow CC)=\frac{N_C}{64\pi}\frac{h^4}{(32\pi^2)^2}\frac{M_{H}^3}{F_\mathrm{S}^2}\left(1-\frac{4M_{C}^2}{M_{H}^2}\right)^{3/2} \,.
\end{equation}
After some order assumptions $N_C h^4\approx100$, and $M_H\sim M_C\sim F_\mathrm{S}\sim \Lambda_\F$
we get an estimate
\begin{equation}
\Gamma(H\rightarrow CC)\approx 10^{-4}\Lambda_\F
\end{equation}
leading to enormously fast decay.

In the version (63333), if it is kinematically allowed, a decay to $N_\mathrm{heavy}$ super-heavy neutrinos of mass
$m_{\mathrm{heavy}}\sim\Lambda_\F$ (that are absent in the version (333)), gives a
contribution to the heavy sterile majoron decay width of comparable size with
\eqref{DW_HM_CC}. From the effective vertex \eqref{L_Hnn} the decay width follows
\begin{equation}\label{DW_H_nn}
\Gamma(H\rightarrow nn)=\frac{N_\mathrm{heavy}}{8\pi}\frac{m_{\mathrm{heavy}}^2}{F_{\mathrm{S}}^2} M_H\left(1-\frac{4m_\mathrm{heavy}^2}{M_{H}^2}\right)^{3/2}
\end{equation}
and a rough estimate is
\begin{equation}
\Gamma(H\rightarrow nn)\approx 10^{-1}\Lambda_\F \,.
\end{equation}

\section{Conclusions}\label{secVI}
The intention of this paper was to investigate the sterile particle sector
of the flavor gauge model \cite{Hosek:2009ys,Hosek:NagoyaProceeding,Benes:2011gi}
of the electroweak symmetry breaking.

The model possesses a nice feature that its consistence requires the existence
of a definite number of right-handed neutrino fields. Together with the left-handed
neutrinos they form Majorana mass eigenstates, what is triggered by the formation
of their self-energies dynamically.

The neutrino self-energies break the global symmetries giving rise to majorons.
We cannot compute any fermion mass spectrum. But if neutrinos acquire Majorana masses
dynamically, majorons must exist. The existence of the standard majoron as a consequence
of the spontaneous lepton number breaking is an inevitable
outcome of all versions of the model. The existence of the set of light sterile
majorons, and one super-heavy sterile majoron depends on whether the sterility symmetry
is broken, and their particular spectrum depends on how it is broken and differs from
version to version. Majorons are both left- and right-handed neutrino composites. If the standard
and the light sterile majorons acquire mass from the gravitational effects, they
are excellent candidates for a warm dark matter \cite{Berezinsky:1993fm,Lattanzi:2008zz}.
The heavy sterile majoron is too unstable to account for any amount of matter of the Universe.

In any case, the heavy sterile majoron provides the Peccei--Quinn mechanism that
eliminates the flavor $\theta$-term from the effective Lagrangian. The heavy
sterile majoron is the composite invisible flavor axion. In this paper we have
ignored anomalies of charged fermion Abelian currents in order to concentrate
only on the neutrino sector as much as possible. The model in its completeness is analyzed
in \cite{Benes:2011gi}. In the simplified case, due to the presence of flavor axion, the heavy sterile majoron, the
model does not suffer
from a $CP$ violation originating from the non-trivial topology of the flavor gauge
dynamics. The only sources of the $CP$ violation remain to be un-transformable
phases of the neutrino mixing matrices in the flavor gauge interactions. The $CP$
violating phases originate from the non-trivial neutrino self-energies $\Sigma$.

The sterile particle spectrum is the first main result of the paper. It is qualitatively
common to all chiral versions of the model. The analysis is, nevertheless, based on the
crucial assumption that the flavor symmetry scenario actually happens.

As the second main result of the paper we brought several heuristic but meaningful
arguments why we see the non-minimal chiral version of the model with sextet right-handed
neutrinos favored. To our surprise, such version has appeared to be also phenomenologically
most suitable.

First of all, better understanding of the flavor symmetry self-breaking has been
reached within the (63333) version. Just in the analogy with the color superconductivity,
at the extremely high energy scale $\Lambda_\mathrm{F}$ the right-handed neutrino
fields form the Majorana condensates that break flavor but not electroweak symmetry.
The right-handed neutrinos and flavor gauge bosons acquire extremely high masses. The
presence of the sextet right-handed neutrino fields is crucial: Their pairing
channels are the most attractive, therefore their condensation happens at the highest
energy scale which is \emph{naturally} separated from the energy scale where the rest
of fermion self-energies are formed and the electroweak symmetry is broken. This lower
scale is, nevertheless, connected to the scale where the QCD axion is formed, i.e.,
$10^9-10^{12}\,\mathrm{GeV}$ \cite{Benes:2011gi}.  So there is no advantage against the version (333) in
explaining the smallness of the charged fermion masses. We still need the huge amplification
of scales. It turns out that the right-handed neutrino
Majorana self-energies must be generated at much higher scale $\Lambda_\mathrm{F}$
than $10^{12}\,\mathrm{GeV}$.

Second, the strongly coupled right-handed neutrino condensate formed at this very high
energy scale is phenomenologically welcome. (i) It can generate the baryogenesis
and drive the inflation of the Universe \cite{Barenboim:2008ds,Barenboim:2010nm}. (ii) It naturally provides
the see-saw pattern of the neutrino mass matrix and suggests $\Lambda_\F\gtrsim10^{14}\,\mathrm{GeV}$.

Third, in order to reach the presence of the three light electroweak neutrinos in
the particle spectrum, we were forced to assume special (but not unnatural) form
of the neutrino mass matrix. The form preserves the residual sterility symmetry
that protects the see-saw mechanism. It also protects smallness of masses of a number of decoupled
sterile neutrinos that can possibly account for fermionic warm dark matter \cite{Nieuwenhuizen:2008pf,Kusenko:2009up,Bezrukov:2009th}.

\begin{acknowledgments}
The author gratefully acknowledges discussions with J. Ho\v{s}ek, G. Barenboim, J. Novotn\'y, and P. Bene\v{s}. The work was supported by the Grant LA08015 of the Ministry of Education of the Czech Republic.
\end{acknowledgments}

\appendix
\section{Nambu--Gorkov formalism}\label{appA}

The neutrino fields are accommodated within the Nambu--Gorkov multispinor $n$ defined in \eqref{NGmultiplet}.
Its canonical anti-commutation relations then follow
\begin{subequations}
\begin{eqnarray}
\{n_{\alpha i}(x),n_{\beta j}^\dag(y)\}_{\mathrm{E.T.}} & = & \delta_{ij}\delta_{\alpha\beta}\delta^{(3)}(\mathbf{x}-\mathbf{y})\,, \\
\{n_{\alpha i}(x),n_{\beta j}(y)\}_{\mathrm{E.T.}} & = & \delta_{ij}[C\gamma_0]_{\alpha\beta}\delta^{(3)}(\mathbf{x}-\mathbf{y})\,,\hspace{0.5cm}
\end{eqnarray}
\end{subequations}
where $C$ is charge conjugation matrix.

The flavor transformations
\begin{equation}\label{n_sterile_transformation}
n'=\mathrm{e}^{\im\theta^a t^a}n \,,
\end{equation}
are generated by the flavor generators
\begin{equation}\label{NGflavorGenerator}
t^a=\beginm{ccc} T_{\mathbf{3}}^{a}P_\mathrm{R}-[T_{\mathbf{3}}^{a}]^\mathrm{T}P_\mathrm{L} & & \\
                            & \hspace{-2cm}\openone_{4\times4}\left(T_{\mathbf{3}}^{a}P_\mathrm{L}-[T_{\mathbf{3}}^{a}]^\mathrm{T}P_\mathrm{R}\right) & \\
                            & & \hspace{-2cm}T_{\mathbf{6}}^{a}P_\mathrm{R}-[T_{\mathbf{6}}^{a}]^\mathrm{T}P_\mathrm{L} \endm\,.\hspace{0.5cm} \nonumber
\end{equation}
The lepton number transformation of the neutrino fields is
\begin{equation}\label{n_sterile_transformation}
n'=\mathrm{e}^{\im\theta l}n \,,
\end{equation}
where $l$ denotes the corresponding generator
\begin{equation}\label{NGleptonNumberGenerator}
l=\beginm{ccc} -L_\mathrm{EW}\gamma_5 &     &   \\
                                           & \hspace{-0.3cm}\frac{1}{4}a\openone_{4\times4}\gamma_5 &   \\
                                           &     & \hspace{-0.3cm}(1-a)\gamma_5 \endm\,.\hspace{0.5cm}
\end{equation}
The sterility transformations of the neutrino fields are
\begin{equation}\label{n_sterile_transformation}
n'=\mathrm{e}^{\im\theta_\alpha s_\alpha}n
\end{equation}
and the corresponding currents of the sterility symmetry are compactly
rewritten as
\begin{equation}
j_{\mathrm{S},\alpha}^\mu = \frac{1}{2}\bar{n}\gamma^\mu s_\alpha n
\,.
\end{equation}
where $s_\alpha$ schematically denotes generators of all
the sterility symmetries
\begin{subequations}\label{NGsterileGenerator}
\begin{eqnarray}
S_3-S_6: & & s_0=\beginm{ccc} 0 &     &   \\
                                           & \frac{1}{4}\openone_{4\times4}\gamma_5 &  \\
                                           &     & -\gamma_5 \endm\,; \nonumber\\
\SU{4}_\mathrm{S}: & & s_i=\beginm{ccc} 0 &     &   \\
                                           & S_{i}P_\mathrm{R}-S_{i}^\mathrm{T}P_\mathrm{L} &   \\
                                           &     & 0 \endm\,, \ i=1,..,15\,; \nonumber\\
S_3+S_6: & & s_{16}=\beginm{ccc} 0 &     &   \\
                                           & \frac{1}{4}\openone_{4\times4}\gamma_5 &   \\
                                           &     & \gamma_5 \endm\,.
\end{eqnarray}
\end{subequations}

\section{Two-loop $\beta$-function}\label{appB}

Two-loop $\beta$-function is given by \cite{Machacek:1983tz}
\begin{eqnarray}\label{beta2}
& & \beta(h) =  \nonumber\\
& & \hspace{0.2cm}-\frac{h^3}{(4\pi)^2}\left[\frac{11}{3}C(8)-\frac{2}{3}N^{\mathrm{EW}}C(3)-\frac{2}{3}\sum_\mathbf{r} N^{\nu_\mathrm{R}}_\mathbf{r} C(\mathbf{r})\right] \nonumber\\
& & \hspace{0.2cm}-\frac{h^5}{(4\pi)^4}\left[\frac{34}{3}C(8)^2-N^{\mathrm{EW}}\left(2C_2(3)+\frac{10}{3}C(8)\right)C(3)
\right. \nonumber\\
& & \hspace{0.2cm}\left.
-\sum_\mathbf{r} N^{\nu_\mathrm{R}}_\mathbf{r}\left(2C_2(\mathbf{r})+\frac{10}{3}C(8)\right)C(\mathbf{r})\right] \,,
\end{eqnarray}
where the coefficient $C(\mathbf{r})$ reflects the flavor symmetry representation of the
right-handed neutrino field, and is related to the quadratic Casimir invariant
$C_2(\mathbf{r})$. Their definitions and their relation are
\begin{subequations}
\begin{eqnarray}
\delta^{ab}C(\mathbf{r}) & = & \Tr{T^{a}_\mathbf{r} T^{b}_\mathbf{r}} \,,\\
d(\mathbf{r})C_2(\mathbf{r}) & = & \Tr{T^{a}_\mathbf{r} T^{a}_\mathbf{r}} \,,\\
d(\mathbf{r})C_2(\mathbf{r})  & = & d(G)C(\mathbf{r}) \,.
\end{eqnarray}
\end{subequations}

\begin{table}[t]
\begin{tabular}{lc|cc|cc}
$\mathbf{r}$ & $d(\mathbf{r})$ & $C(\mathbf{r})$ & $C_2(\mathbf{r})$ & $A(\mathbf{r})$ & $C_3(\mathbf{r})$ \\
\hline
\hline
$\mathbf{3}(\overline{\mathbf{3}})$   & $3$      & $1/2$  & $4/3$  & $(-)1$  & $(-)10/9$ \\
$\mathbf{6}(\overline{\mathbf{6}})$   & $6$      & $5/2$  & $10/3$ & $(-)7$  & $(-)35/9$ \\
$\mathbf{8}$                          & $8$ & $3$    & $3$    & $0$     & $0$       \\
$\mathbf{10}(\overline{\mathbf{10}})$ & $10$     & $15/2$ & $6$    & $(-)27$ & $(-)9$ \\
\hline
\hline
\end{tabular}
\caption{\small List of important coefficients for the lowest representations of the group $\SU{3}$. }
\label{table}
\end{table}

For completeness we mention also the anomaly coefficient $A(\mathbf{r})$ important for the
anomaly analysis. It is related to the cubic Casimir invariant $C_3(\mathbf{r})$. The relevant formulas are
\begin{subequations}
\begin{eqnarray}
\frac{1}{2}d^{abc}A(\mathbf{r}) & = & \Tr{T^{a}_\mathbf{r}\{T^{b}_\mathbf{r},T^{c}_\mathbf{r}\}} \,, \label{anomalyC}\\
d(\mathbf{r}) \, C_3(\mathbf{r}) & = & d^{abc}\Tr{T^{a}_\mathbf{r}\,T^{b}_\mathbf{r}\,T^{c}_\mathbf{r}} \,, \label{CasimirI}\\
2 d(\mathbf{r}) \, C_3(\mathbf{r})  & = & \frac{5}{6}\,d(G)\,A(\mathbf{r}) \,.
\end{eqnarray}
\end{subequations}
The values for some of the lowest representations are listed in Tab.~\ref{table}.

\section{Flavor gauge boson mass matrices}\label{appC}

\begin{widetext}
\begin{subequations}\label{Mgauge}
\begin{eqnarray}
M_{6}^2 & = & h^2
\beginm{cccccccc}
(\phi_1+\phi_2)^2 \hspace{-7pt}& 0 & 0 & 0 & 0 & 0 & 0 & 0 \\
0 & \hspace{-7pt} (\phi_1-\phi_2)^2  \hspace{-7pt}& 0 & 0 & 0 & 0 & 0 & 0 \\
0 & 0 & \hspace{-7pt} 2(\phi_{1}^2+\phi_{2}^2)  \hspace{-7pt}& 0 & 0 & 0 & 0 & \frac{2}{\sqrt{3}}(\phi_{1}^2-\phi_{2}^2) \\
0 & 0 & 0 & \hspace{-7pt} (\phi_1+\phi_3)^2  \hspace{-7pt}& 0 & 0 & 0 & 0 \\
0 & 0 & 0 & 0 & \hspace{-7pt} (\phi_1-\phi_3)^2  \hspace{-7pt}& 0 & 0 & 0 \\
0 & 0 & 0 & 0 & 0 & \hspace{-7pt} (\phi_2+\phi_3)^2  \hspace{-7pt}& 0 & 0 \\
0 & 0 & 0 & 0 & 0 & 0 & \hspace{-7pt} (\phi_2-\phi_3)^2  \hspace{-7pt}& 0 \\
0 & 0 & \hspace{-7pt} \frac{2}{\sqrt{3}}(\phi_{1}^2-\phi_{2}^2) \hspace{-7pt} & 0 & 0 & 0 & 0 & \frac{2}{3}(\phi_{1}^2+\phi_{2}^2+4\phi_{3}^2)
\endm\,,\\
\nonumber\\
M_{3}^2 & = & \frac{h^2}{4}
\beginm{cccccccc}
(\varphi_{4}^2+\varphi_{5}^2) \hspace{-10pt} & 0 & 0 & \varphi_{5}\varphi_{6} & 0 & \varphi_{4}\varphi_{6} & 0 & \frac{2}{\sqrt{3}}\varphi_{4}\varphi_{5} \\
0 & \hspace{-10pt} (\varphi_{4}^2+\varphi_{5}^2) \hspace{-10pt} & 0 & 0 & \varphi_{5}\varphi_{6} & 0 & -\varphi_{4}\varphi_{6} & 0 \\
0 & 0 & \hspace{-10pt} (\varphi_{4}^2+\varphi_{5}^2) \hspace{-10pt} & \varphi_{4}\varphi_{6} & 0 & -\varphi_{5}\varphi_{6} & 0 & \frac{1}{\sqrt{3}}(\varphi_{4}^2-\varphi_{5}^2) \\
\varphi_{5}\varphi_{6} & 0 & \varphi_{4}\varphi_{6} & \hspace{-10pt} (\varphi_{4}^2+\varphi_{6}^2) \hspace{-10pt} & 0 & \varphi_{4}\varphi_{5} & 0 & -\frac{1}{\sqrt{3}}\varphi_{4}\varphi_{6} \\
0 & \varphi_{5}\varphi_{6} & 0 & 0 & \hspace{-10pt} (\varphi_{4}^2+\varphi_{6}^2) \hspace{-10pt} & 0 & \varphi_{4}\varphi_{5} & 0 \\
\varphi_{4}\varphi_{6} & 0 & -\varphi_{5}\varphi_{6} & \varphi_{4}\varphi_{5} & 0 & \hspace{-10pt} (\varphi_{5}^2+\varphi_{6}^2) \hspace{-10pt} & 0 & -\frac{1}{\sqrt{3}}\varphi_{5}\varphi_{6} \\
0 & -\varphi_{4}\varphi_{6} & 0 & 0 & \varphi_{4}\varphi_{5} & 0 & \hspace{-10pt} (\varphi_{5}^2+\varphi_{6}^2) \hspace{-10pt} & 0 \\
\frac{2}{\sqrt{3}}\varphi_{4}\varphi_{5} & 0 &  \hspace{-10pt}\frac{1}{\sqrt{3}}(\varphi_{4}^2-\varphi_{5}^2) & -\frac{1}{\sqrt{3}}\varphi_{4}\varphi_{6} & 0 & -\frac{1}{\sqrt{3}}\varphi_{5}\varphi_{6} & 0 & \hspace{-10pt} \frac{1}{3}(\varphi_{4}^2+\varphi_{5}^2+4\varphi_{6}^2)
\endm\,.
\end{eqnarray}
\end{subequations}
\end{widetext}

\bibliography{references}

\end{document}